%% file: 3-DHNM.tex
\documentclass{pj}
\usepackage{amssymb}
\usepackage{graphics}
\usepackage{latexsym,amsmath,amsfonts,amscd,subfigure}
\usepackage{epstopdf}
\usepackage{epsfig}
\usepackage{changebar}
\usepackage{indentfirst}
\usepackage{verbatim}
\usepackage{graphicx}
\usepackage{amsmath,amsthm}
\usepackage{extarrows}
\input latex_defs.tex

\begin{document}
\setcounter{page}{1}
\pjheader{Vol.\ x, y--z, 2026}

\title[A 3-D HNM for Simulating EM Fields in Structures with Multiple Inhomogeneous Layered Media]
{A 3-D Hybrid Numerical Method for Simulating Electromagnetic Fields in Structures with Multiple Inhomogeneous Layered Media}
 \footnote{\it Received date}
 \footnote{\hskip-0.12in*\, Corresponding
author: Qing Huo Liu (qhliu@eitech.edu.cn).}
\footnote{\hskip-0.12in\textsuperscript{1} School of Mathematical Sciences, Guizhou Normal University, Guiyang 550025, China. \\
\textsuperscript{2} Institute of Electromagnetics and Acoustics, Xiamen University, Xiamen 361005, China.\\
\textsuperscript{3} Eastern Institute of Technology, Ningbo 315200, China.}

\author{Jie Liu\textsuperscript{1,3}, Taihe Li\textsuperscript{2}, Ke Chen\textsuperscript{2}, Mingwei Zhuang \textsuperscript{2} and Qing Huo Liu\textsuperscript{*,3,2}}

\runningauthor{Jie Liu, Taihe Li, Ke Chen, Mingwei Zhuang and Qing Huo Liu}

\tocauthor{Institution of Electromagnetics and Acoustics, Xiamen, 361005, P.R. China.  and FistName1~LastName1}

\begin{abstract}
A three-dimensional (3-D) hybrid numerical method (HNM) is presented for electromagnetic scattering in structures with multiple inhomogeneous layered media coupled to an arbitrary 3-D non-layered scattering region. It integrates the 3-D numerical mode-matching (NMM) method with a tree-cotree-based mixed finite element method (MFEM). In the NMM, the fields in the 3-D layered media are reduced to a superposition of 2-D waveguide eigenmodes, while the MFEM discretizes the 3-D scattering region. The HNM thus inherits the dimensionality-reduction advantages of both conventional hybrid MM/FEM and pure NMM. Numerical experiments show that the HNM provides an efficient alternative for scattering problems involving non-layered regions embedded in layered media.
\end{abstract}

\noindent{\bf\small Keywords~:~\scriptsize}{\small Electromagnetic (EM) scattering, inhomogeneous layered media, 3-D mixed finite element method (MFEM), numerical mode-matching (NMM) method, hybrid numerical method.}

\setlength {\abovedisplayskip} {6pt plus 3.0pt minus 4.0pt}
\setlength {\belowdisplayskip} {6pt plus 3.0pt minus 4.0pt}

\section{Introduction}
\label{section label}

Efficient and accurate simulation of electromagnetic (EM) radiation and scattering is essential in a variety of fields, such as remote sensing, antenna theory, geophysical exploration, biomedical imaging, interconnects, microwave integrated circuits, and so on\cite{Kahnert2003,Zhdanov2002,Abubakar2002,Kinayman1997,Hu2001}.
As structures become increasingly complex, involving curvature, corners, apertures, and inhomogeneous dielectric loading, the analytical approaches become intractable, driving the development of numerical methods\cite{Umashankar1982}.
For EM scattering simulations, the finite-difference time-domain (FDTD) method \cite{Yee1997}, integral equation methods (surface \cite{Graglia1989} and volume \cite{Arrieta2022,Cao2017} formulations), and the finite element method (FEM) \cite{Jin1993b,Kirsch2002,Harmon2021} have each demonstrated effectiveness. However, as models grow in complexity, single-method approaches incur prohibitive computational costs, motivating hybrid techniques that combine method-specific advantages.

Among hybrid approaches, several hybrid mode-matching/finite element methods (MM/FEM) have been developed for cascaded waveguide structures \cite{Beyer1995, Arndt1997, Arndt2004, Omar1985, Arena2000, Wakasa2019,Rubio1999}. These methods reduce computational dimensionality by applying mode-matching to regular waveguide sections. While effective in many cases, these existing formulations can be further improved to broaden their range of applicability. For instance, some formulations are based on TE/TM modes and rely on mode orthogonality \cite{Beyer1995, Arndt1997, Arndt2004, Omar1985, Wakasa2019}, which may limit their use in inhomogeneous or anisotropic waveguides where hybrid modes are present. In other cases, the enforcement of Gauss's law for the divergence-free conditions is not explicitly addressed \cite{Arena2000,Rubio1999}; as a result, when 2-D FEM is used for eigenmode solving, spurious DC modes may appear \cite{Liu2020feb2}, and when 3-D FEM is employed to discretize the scattering region, potential accuracy degradation or numerical instability has been reported \cite{Chen2020}. Additionally, the derivation of the generalized scattering matrix (GSM) via the generalized admittance matrix (GAM) involves a system matrix inversion \cite{Rubio1999}, which may become costly for large-scale problems. These observations motivate the development of a hybrid approach that combines the strengths of the numerical mode-matching (NMM) method and tree-cotree-based mixed finite element method (MFEM) while addressing the above aspects.

\begin{figure}[h]
  \centering
  {
   \includegraphics[width=0.6\columnwidth,draft=false]{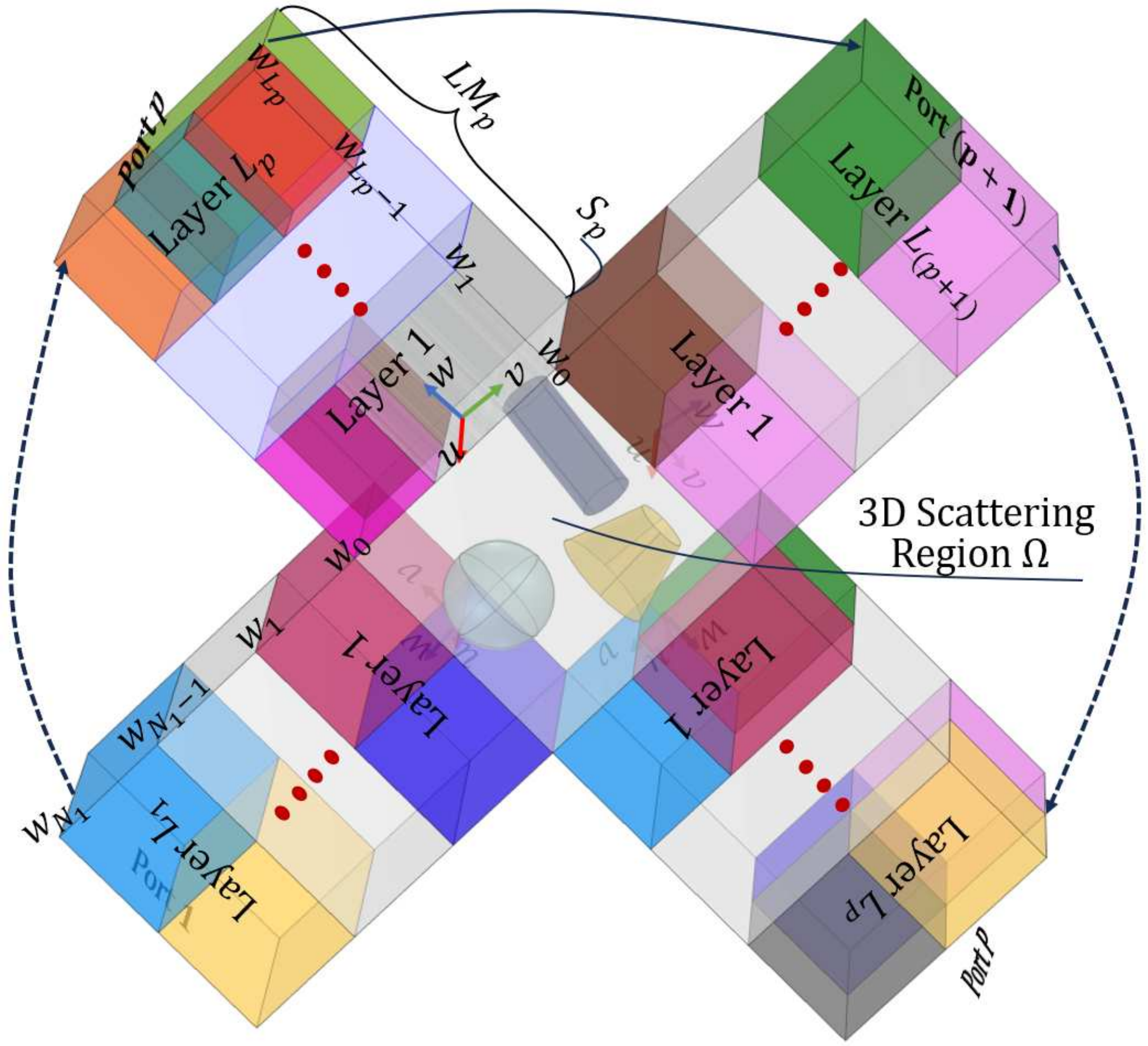}
   }
 \caption{Sketch of EM scattering by a non-layered 3-D scattering region $\Omega$ with the multiple scatterers. The region $\Omega$ connects to multiple layered waveguides $LM_{p}$ ($p=1,2,\ldots P$) via interface $S_p$ at $w=w_{0}$. Each $LM_{p}$ is divided into $L_p$ layers along the $w$-direction at interfaces $w=w_{\alpha}$ ($\alpha=0,1,\ldots,L_{p}$), with the
$p$-th port at the end of the $L_{p}$-th layer. Layer counts may differ among $LM_{p}$, and each layer is filled with inhomogeneous media in the local $u$-and $v$-directions. Outer boundaries can be PEC, PMC or Bloch periodic, except at the ports.
An incident wave excites the scatterers within $\Omega$ through $LM_{p}$ via the $p$-th port.}
\label{sketch0}
\end{figure}

The NMM method offers an alternative semi-analytical approach that reduces 3-D problems to 2-D eigenvalue problems and 1-D layered media problems, solvable via recursion \cite{Pudensi1982}. The NMM method has been successfully applied to multilayered transmission lines \cite{Kamra2019}, orthogonal-plano-cylindrically layered structures \cite{Dai2015}, optical fibers with Kerr nonlinearity \cite{Wu2023}, metasurfaces \cite{Liu2019,Liu2020feb1,Liu2020feb2,Tong2022}, EM well logging \cite{Liu2021}, and fully anisotropic and nonreciprocal structures\cite{Liu2022july}. However, pure NMM cannot handle arbitrary 3-D scattering regions that are not amenable to layered representation along a single direction, as shown in Fig. \ref{sketch0}.

To address these challenges, a hybrid numerical method (HNM) is developed that uniquely combines the 3-D NMM and the 3-D tree-cotree-based MFEM \cite{Venkatarayalu2006,Chen2020,Wang2021}.
This synthesis offers three main advantages over existing approaches.

1) The NMM component employs a spurious-free 2.5-D MFEM to compute eigenmodes of arbitrarily shaped inhomogeneous anisotropic waveguides under various boundary conditions (PEC, PMC, Bloch periodic), where the 2.5-D MFEM incorporates Gauss's law. Generalized reflection matrices are derived recursively using local reflection, transmission, and propagation matrices, independent of orthogonal eigenmods, providing semi-analytical solutions for the layered media with substantially reduced computational cost.

2) The 3-D MFEM component uses tree-cotree decomposition to discretize the scattering region, where the electric field is decomposed into cotree-edge and tree-edge components, with the latter represented by gradients of nodal basis functions. This approach enforces both tangential continuity and the material-weighted divergence-free condition (i.e., $\nabla\cdot \bar{\bar{\epsilon}}_r \textbf{E} =0$) in the discrete sense, without invoking Gauss' law via Lagrange multipliers \cite{Fumio1987}, thus avoiding additional unknowns.

3) The tangential continuity of EM fields is enforced as exact boundary conditions at the interfaces $S_p$ between $LM_p$ and $\Omega$, eliminating numerical errors from approximate numerical fluxes.

The HNM inherits the dimensionality-reduction advantage of both conventional hybrid MM/FEM and the pure NMM, while improving upon certain aspects of the hybrid MM/FEM where further development is desirable. Beyond our earlier works \cite{Liu2019,Liu2020feb1,Liu2020feb2,Tong2022,Liu2021,Liu2022july}, which employed the 2.5-D MFEM/MSEM only as an eigenmode solver within the pure NMM framework, the proposed method advances to a fully hybrid formulation by integrating the 3-D NMM with the tree-cotree MFEM. This integration also addresses the gap left by the standalone 3-D tree-cotree MFEM in \cite{Chen2020} and \cite{Wang2021}, which, though effective for eigenvalue and time-domain problems, was not designed for coupling with layered media or ports. Consequently, the HNM enables simulations of arbitrary 3-D scattering regions connected to multiple inhomogeneous layered media in a unified variational framework.

The article is organized as follows. Section 2 presents the formulations of the hybrid HNM; Section 3 demonstrates accuracy and efficiency through five numerical examples, including multi-layered media with scatterers, bent waveguides, multiple scatterers, a millimeter-wave circulator model, and a three-port divider model; Section 4 concludes. The bilinear functions used in the variational forms and the detailed elements of the system matrices are provided in Appendices A, B, and C, respectively.

\section{Formulation}
\label{sec:formulation}
\begin{figure}[h]
\centering
{
\includegraphics[width=0.6\columnwidth,draft=false]{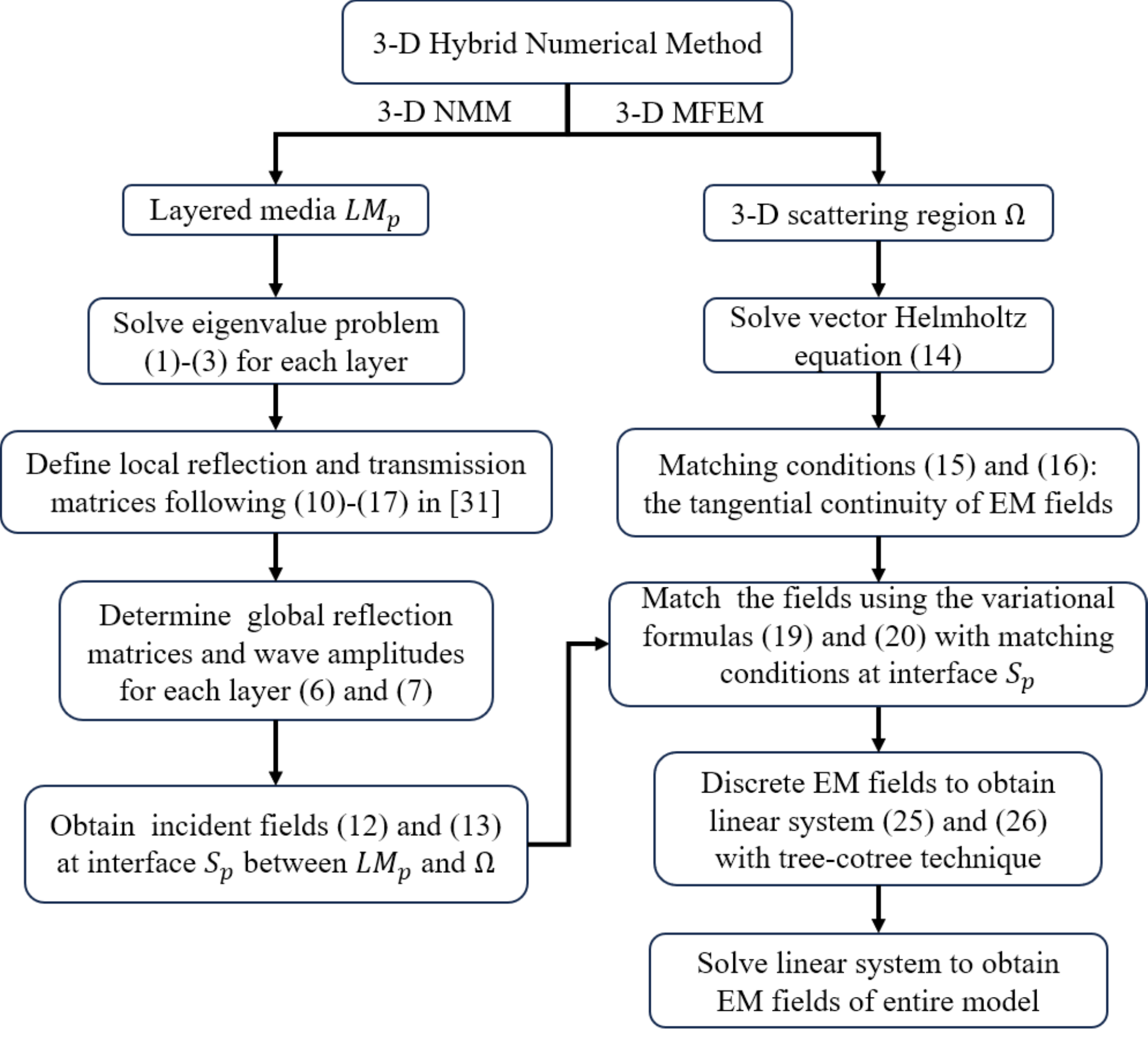}
}
\caption{Flowchart of the 3-D hybrid numerical method (HNM).}
\label{sketch1}
\end{figure}

The hybrid numerical method (HNM) outlined in Fig. \ref{sketch1} is used to simulate the model shown in Fig. \ref{sketch0}, which consists of inhomogeneous layered media $LM_{p}$ ($p=1,2,\ldots,P$) and a 3-D scattering region $\Omega$. The 3-D MFEM with tree-cotree technique is employed to discretize $\Omega$,
while each $LM_p$ is treated as a layered waveguide along the $w$-direction, enabling the 3-D NMM method to describe the EM fields within these waveguides.

\subsection{NMM Method for the Inhomogeneous Layered Media}

A key step in the NMM method is to obtain the waveguide eigenmodes for each layer within the inhomogeneous layered media $LM_p$.
Thus, we start by solving the eigenmodes in an inhomogeneous waveguide. In this article, $\textbf{E}$ and $\textbf{H}$ denote the electric and magnetic
fields, respectively. The notations for scalars, vectors, and matrices are summarized in Table \ref{example00}.

\begin{table}[h]
\renewcommand{\arraystretch}{1.5}
\caption{Notation of Scalar, Vector and Matrix}
\centering
\begin{tabular}{cccccc}
\hline
Quantity   & Scalar   & Column Vector & Matrix & Column Matrix\\
\hline
Notation    & v/$\epsilon$    & $\textbf{v}$/$\bf{\Phi}$ & $\bar{\bar{\epsilon}}$ / $\bar{\bar{A}}$ & $\bar{\bar{\textbf{A}}}$\\
\hline
\end{tabular}
\label{example00}
\end{table}

\emph{1) Waveguide eigenvalue problem for each layer:} To obtain the eigenmodes in $LM_p$,
each individual layer $n$ ($n=1,2,\ldots, L_p$) is treated as a waveguide along the $+w$-direction in the local coordinates system ($u,v,w$).
The governing equations for an inhomogeneous anisotropic waveguide are given by \cite{Liu2020feb2}
\begin{subequations}\label{eq:1}
\begin{align}
\nabla_{t}\times\mu_{rw}^{(n)-1}\nabla_{t}\times\textbf{e}_{t}^{(n)}+\bar{\bar{\mathfrak{R}}}\bar{\bar{\mu}}_{rt}^{-1}\bar{\bar{\mathfrak{R}}}\nabla_{t}e_{w}^{new}
-k_{0}^{2}\bar{\bar{\epsilon}}_{rt}^{(n)}\textbf{e}_{t}^{(n)}
=(k_{w}^{(n)})^{2}\bar{\bar{\mathfrak{R}}}\bar{\bar{\mu}}_{rt}^{(n)-1}\bar{\bar{\mathfrak{R}}}\textbf{e}_{t}^{(n)}\\
\nabla_{t}\cdot(\bar{\bar{\epsilon}}_{rt}^{(n)}\textbf{e}_{t}^{(n)})-\epsilon_{rw}^{(n)}e_{w}^{new}=0.
\end{align}
\end{subequations}
Here, $k_{0}$ is the free-space wavenumber; $\bar{\bar{\epsilon}}_{rt}^{(n)}$ and $\bar{\bar{\mu}}_{rt}^{(n)}$ are the transverse relative permittivity and permeability tensors, respectively, $\epsilon_{rw}^{(n)}$ and $\mu_{rw}^{(n)}$ are longitudinal components, all independent of $w$; $\textbf{e}_{t}^{(n)}$ and $e_{w}^{new}=jk_{w}^{(n)}e_{w}^{(n)}$ are the transverse and $w$ components, respectively. $k_{w}^{(n)}$ is the propagation constant along the $+w$-direction; $\bar{\bar{\mathfrak{R}}}=[\begin{smallmatrix} 0 &-1\\1&0\end{smallmatrix}]$ is the rotation matrix equivalent to $\hat{w}\times$. The outer boundaries of $LM_p$ may be PEC, PMC, or Bloch periodic. The eigenmodes are obtained by solving (\ref{eq:1}) using the 2.5-D MFEM, which eliminates all spurious modes present in the conventional FEM \cite{Liu2020feb2}. The transverse and longitudinal magnetic field components are then obtained from
\begin{equation}\label{eq:2}
\mu_{rw}^{(n)-1}(\nabla_{t}\times\textbf{e}_{t}^{(n)})=-j\omega\mu_{0}h_{w}^{(n)}\hat{w}
\end{equation}
\begin{equation}\label{eq:3}
-\bar{\bar{\mu}}_{rt}(\bar{\bar{\mathfrak{R}}}\nabla_{t}e_{w}^{(n)}+jk_{w}^{(n)}\bar{\bar{\mathfrak{R}}}\textbf{e}_{t}^{(n)})=-j\omega\mu_{0}\textbf{h}_{t}^{(n)}
\end{equation}
where $\mu_{0}$ is permeability in vacuum and $\omega$ denotes the angular frequency.

\emph{2) EM fields in each layer:} Once the eigenmodes of the $n$-th layer in $LM_p$ are obtained from (\ref{eq:1})-(\ref{eq:3}), as described in \cite{Liu2022july}, the transverse EM fields for a wave propagating along the $+w$-direction are expressed as
\begin{equation}\label{eq:4}
\begin{split}
\textbf{E}_{n,t}=[(\bar{\bar{\textbf{F}}}_{n}^{+})^{t}e^{-j\bar{\bar{K}}_{n,w}^{+}(w-w_{n-1})}
+(\bar{\bar{\textbf{F}}}_{n}^{-})^{t}e^{-j\bar{\bar{K}}_{n,w}^{-}(w-w_{n})}
\tilde{\bar{G}}_{n,n+1}\bar{\bar{P}}_{n}^{+}]\textbf{A}_{n}
\end{split}
\end{equation}
\begin{equation}\label{eq:5}
\begin{split}
\bar{\bar{\mathfrak{R}}}\textbf{H}_{n,t}=[(\bar{\bar{\textbf{H}}}_{n}^{+})^{t}e^{-j\bar{\bar{K}}_{n,w}^{+}(w-w_{n-1})}
+(\bar{\bar{\textbf{H}}}_{n}^{-})^{t}e^{-j\bar{\bar{K}}_{n,w}^{-}(w-w_{n})}
\tilde{\bar{G}}_{n,n+1}\bar{\bar{P}}_{n}^{+}]\textbf{A}_{n}
\end{split}
\end{equation}
where $\bar{\bar{P}}_{n}^{+}$ is the $+w$-going propagator matrix from interfaces $w_{n-1}$ to $w_n$.
Meanwhile, the matrices $(\bar{\bar{\textbf{F}}}_{n}^{\pm})^{t}$, $(\bar{\bar{\textbf{H}}}_{n}^{\pm})^{t}$ and $\bar{\bar{K}}_{n,w}^{\pm}$
are constructed from $m$ eigenmodes $(k_{\alpha,w}^{\pm(n)},\textbf{e}_{\alpha,t}^{\pm(n)},\textbf{h}_{\alpha,t}^{\pm(n)})$, $\alpha=1,2,\ldots,m$. $\tilde{\bar{G}}_{n,n+1}$ is the global reflection matrix and $\textbf{A}_{n}$ is the amplitude of the $+w$-going wave, which are obtained from the following recursive relations
\begin{equation}\label{eq:6}
\begin{split}
\tilde{\bar{G}}_{n,n+1}=\bar{\bar{R}}_{n,n+1}+\bar{\bar{T}}_{n+1,n}\bar{\bar{P}}_{n+1}^{-}
\times\tilde{\bar{G}}_{n+1,n+2}\bar{\bar{P}}_{n+1}^{+}\bar{\bar{M}}_{n+1}^{(+)}\bar{\bar{T}}_{n,n+1}.
\end{split}
\end{equation}
\begin{equation}\label{eq:7}
\begin{split}
\textbf{A}_{n}=\bar{\bar{M}}_{n}^{(+)}\bar{\bar{T}}_{n-1,n}
\bar{\bar{P}}_{n-1}^{+}\textbf{A}_{n-1}
\end{split}
\end{equation}
where $\bar{\bar{M}}_{n}^{(+)}=[\bar{\bar{I}}-\bar{\bar{R}}_{n,n-1}\bar{\bar{P}}_{n}^{-}\tilde{\bar{G}}_{n,n+1}\bar{\bar{P}}_{n}^{+}]^{-1}$, $\bar{\bar{P}}_{n}^{-}$ is the $-w$-going propagator matrix. The local reflection and transmission matrices $\bar{\bar{R}}_{n,n+1}$, $\bar{\bar{R}}_{n+1,n}$, $\bar{\bar{T}}_{n,n+1}$, and $\bar{\bar{T}}_{n+1,n}$ are obtained from Appendix C of \cite{Liu2022july}.

For the $-w$-going wave, the transverse fields are
\begin{equation}\label{eq:8}
\begin{split}
\textbf{E}_{n,t}=[(\bar{\bar{\textbf{F}}}_{n}^{-})^{t}e^{j\bar{\bar{K}}_{n,w}^{-}(w_{n}-w)}
+(\bar{\bar{\textbf{F}}}_{n}^{+})^{t}e^{j\bar{\bar{K}}_{n,w}^{+}(w_{n-1}-w)}
\tilde{\bar{G}}_{n,n-1}\bar{\bar{P}}_{n}^{-}]\textbf{B}_{n}
\end{split}
\end{equation}
\begin{equation}\label{eq:9}
\begin{split}
\bar{\bar{\mathfrak{R}}}\textbf{H}_{n,t}=[(\bar{\bar{\textbf{H}}}_{n}^{-})^{t}e^{j\bar{\bar{K}}_{n,w}^{-}(w_{n}-w)}
+(\bar{\bar{\textbf{H}}}_{n}^{+})^{t}e^{j\bar{\bar{K}}_{n,w}^{+}(w_{n-1}-w)}
\tilde{\bar{G}}_{n,n-1}\bar{\bar{P}}_{n}^{-}]\textbf{B}_{n}
\end{split}
\end{equation}
where $\textbf{B}_{n}$ denotes the amplitude of the $-w$-going wave and $\tilde{\bar{G}}_{n,n-1}$ is the corresponding global reflection matrix, obtained by recursive relations similar to (\ref{eq:6}) and (\ref{eq:7}) \cite{Liu2022july}.

\emph{3) Incident and scattered fields in layer $1$:} When the $p$-th port is activated by eigenmodes or an incident wave, a wave propagates from layer $L_{p}$ to layer $1$ in $LM_{p}$. From (\ref{eq:4}) and (\ref{eq:5}), the scattered transverse fields in layer $1$ can be written as
\begin{equation}\label{eq:10}
\begin{split}
\textbf{E}_{1,t;p}^{\textrm{scat}}=[(\bar{\bar{\textbf{F}}}_{1,p}^{+})^{t}e^{-j\bar{\bar{K}}_{1,w,p}^{+}(w-w_{0})}
+(\bar{\bar{\textbf{F}}}_{1,p}^{-})^{t}e^{-j\bar{\bar{K}}_{1,w,p}^{-}(w-w_{1})}
\tilde{\bar{G}}_{1,2}^{(p)}\bar{\bar{P}}_{1,p}^{+}]\textbf{A}_{1,p}^{\textrm{scat}}
\end{split}
\end{equation}
\begin{equation}\label{eq:11}
\begin{split}
\bar{\bar{\mathfrak{R}}}\textbf{H}_{1,t;p}^{\textrm{scat}}=[(\bar{\bar{\textbf{H}}}_{1,p}^{+})^{t}e^{-j\bar{\bar{K}}_{1,w,p}^{+}(w-w_{0})}
+(\bar{\bar{\textbf{H}}}_{1,p}^{-})^{t}e^{-j\bar{\bar{K}}_{1,w,p}^{-}(w-w_{1})}
\tilde{\bar{G}}_{1,2}^{(p)}\bar{\bar{P}}_{1,p}^{+}]\textbf{A}_{1,p}^{\textrm{scat}}.
\end{split}
\end{equation}
On the other hand, from (\ref{eq:8}) and (\ref{eq:9}), the incident transverse fields can be expressed as
\begin{equation}\label{eq:12}
\begin{split}
\textbf{E}_{1,t;p}^{\textrm{inc}}=[(\bar{\bar{\textbf{F}}}_{1,p}^{-})^{t}e^{j\bar{\bar{K}}_{1,w,p}^{-}(w_{1}-w)}]\textbf{B}_{1,p}^{\textrm{inc}}
\end{split}
\end{equation}
\begin{equation}\label{eq:13}
\begin{split}
\bar{\bar{\mathfrak{R}}}\textbf{H}_{1,t;p}^{\textrm{inc}}=[(\bar{\bar{\textbf{H}}}_{1,p}^{-})^{t}e^{j\bar{\bar{K}}_{1,w,p}^{-}(w_{1}-w)}]\textbf{B}_{1,p}^{\textrm{inc}}
\end{split}
\end{equation}
where $\textbf{B}_{1,p}^{\textrm{inc}}=\tilde{\bar{T}}_{L_p,1}\textbf{B}_{L_p}$ with the initial condition $\textbf{B}_{L_p}=\textbf{F}_{s}$, and $\textbf{F}_{s}$ is the excitation vector; $\tilde{\bar{T}}_{L_p,1}$ is the global transmission matrix from layer $L_p$ to layer $1$. If the $p$-th port is not excited, only the scattered fields exist in $LM_{p}$. The longitudinal field components are obtained from [31, Eqs. (25)-(32)].

\subsection{Matching Fields between Layered Media and 3-D Scattering Region}

Once the incident and reflected waves in $LM_{p}$ are obtained from (\ref{eq:10})-(\ref{eq:13}), they are coupled with the fields in the 3-D scattering region $\Omega$ through the following three steps. First, the electric field in $\Omega$ is expressed using the vector Helmholtz equation. Second, the boundary conditions at the interface $S_p$ between $LM_{p}$ and $\Omega$ are enforced. Finally, the fields in the two regions are coupled via the variational formulations incorporating these conditions.

\emph{1) Electric field in 3-D scattering region $\Omega$:} The electric field in $\Omega$ satisfies the vector Helmholtz equation
\begin{equation}\label{eq:14}
\begin{split}
\nabla\times\bar{\bar{\mu}}_{r}^{-1}\nabla\times\textbf{E}-k_{0}^{2}\bar{\bar{\epsilon}}_{r}\textbf{E}=\textbf{S}, \textrm{in}~\Omega
\end{split}
\end{equation}
where $\bar{\bar{\epsilon}}_{r}$ and $\bar{\bar{\mu}}_{r}$ are $3\times3$ tensors for the relative permittivity and permeability, respectively; $\textbf{S}$ is the source vector.
To derive the variational formulation of (\ref{eq:14}), we take the inner product with a test function
$\textbf{v}\in \mathbb{H}(\textrm{curl},\Omega)$ and apply integration by parts. The resulting weak form is given by
\begin{equation}\label{eq:15}
\begin{split}
\int_{\Omega}\nabla\times\textbf{v}^{*}\cdot\bar{\bar{\mu}}_{r}^{-1}\nabla\times\textbf{E}dv - k_{0}^{2}\int_{\Omega}\textbf{v}^{*}\cdot\bar{\bar{\epsilon}}_{r}\textbf{E}dv + \\ \int_{\partial\Omega}\textbf{v}^{*}\cdot\hat{n}\times\bar{\bar{\mu}}_{r}^{-1}\nabla\times\textbf{E}d s = \int_{\Omega} \textbf{v}^{*}\cdot \textbf{S} d v
\end{split}
\end{equation}
where $\mathbb{H}(\textrm{curl},\Omega)$ is the curl-conforming space \cite{Hiptmair2002}.

\emph{2) Matching conditions at interface $S_{p}$:}
To match the EM fields between $LM_{p}$ and ${\Omega}$, we need the boundary conditions associated with the tangential continuity of EM fields at the interface $S_p$, that is
\begin{equation}\label{eq:16}
\begin{split}
\hat{w}\times\textbf{H}_{1,t;p}^{\textrm{inc}} + \hat{w}\times\textbf{H}_{1,t;p}^{\textrm{scat}} = \hat{w}\times\textbf{H}_{|_{S_p}}
\end{split}
\end{equation}
\begin{equation}\label{eq:17}
\hat{w}\times\textbf{E}_{1,t;p}^{\textrm{inc}} + \hat{w}\times\textbf{E}_{1,t;p}^{\textrm{scat}} = \hat{w}\times\textbf{E}_{|_{S_p}}
\end{equation}
where $\textbf{H}_{1,t;p}^{\textrm{inc}}$ and $\textbf{H}_{1,t;p}^{\textrm{scat}}$ are the incident and scattered transverse magnetic fields in layer $1$ of $LM_{p}$, respectively. $\textbf{E}_{1,t;p}^{\textrm{inc}}$ and $\textbf{E}_{1,t;p}^{\textrm{scat}}$ are the corresponding transverse electric fields.
$\hat{w}\times\textbf{E}_{|_{S_p}}$ and $\hat{w}\times\textbf{H}_{|_{S_p}}$ denote the tangential EM fields evaluated on the interface $S_p$ from the 3-D MFEM solution in $\Omega$, respectively.

\emph{3) Matching fields:}
The weak form (\ref{eq:15}) is rewritten as the bilinear form: find $\textbf{E}\in\mathbb{H}(\textrm{curl},\Omega)$ and $\textbf{H}\in\mathbb{H}(\textrm{curl},\Omega)$ such that
\begin{equation}\label{eq:18}
\begin{split}
c(\textbf{E},\textbf{v})&-k_{0}^{2}a(\textbf{E},\textbf{v})+s_{0}(\textbf{E},\textbf{v})
-jk_{0}\eta_{0}\sum_{p=1}^{P}s_{p}(\textbf{H},\textbf{v})\\
&=(\textbf{S},\textbf{v}), \forall \textbf{v}\in  \mathbb{H}(\textrm{curl},\Omega)
\end{split}
\end{equation}
where the boundary integral $s_{p}$ is handled using the identity $\hat{w}\times\bar{\bar{\mu}}_{r}^{-1}\nabla\times\textbf{E}=-jk_{0}\eta_{0}\hat{w}\times \textbf{H}$, coupling the electromagnetic fields at the interface $S_{p}$ through the matching conditions; $P$ is the number of ports and $\eta_{0}$ is the wave impedance in vacuum.

Since the boundary integral $s_p$ in (\ref{eq:18}) contains the tangential magnetic field, substituting (\ref{eq:16}) into (\ref{eq:18}) yields

\begin{equation}\label{eq:19}
\begin{split}
s_{p}(\textbf{H},\textbf{v})= s_{p}(\textbf{H}_{1,t;p}^{\textrm{inc}},\textbf{v}) + s_{p}(\textbf{H}_{1,t;p}^{\textrm{scat}},\textbf{v})
\end{split}
\end{equation}
and (\ref{eq:18}) is rewritten as
\begin{equation}\label{eq:20}
\begin{split}
c(\textbf{E},\textbf{v})&-k_{0}^{2}a(\textbf{E},\textbf{v})+s_{0}(\textbf{E},\textbf{v})
-c_{0}\sum_{p=1}^{P}s_{p}(\textbf{H}_{1,t;p}^{\textrm{scat}},\textbf{v})\\
&=(\textbf{S},\textbf{v}) + c_{0}\sum_{p=1}^{P}s_{p}(\textbf{H}_{1,t;p}^{\textrm{inc}},\textbf{v})
\end{split}
\end{equation}
where $c_{0}=jk_{0}\eta_{0}$.
From (\ref{eq:11}) and (\ref{eq:20}), there are the unknowns $\textbf{E}$ and $\textbf{A}_{1,p}^{\textrm{scat}}$ in (\ref{eq:20}). We next employ (\ref{eq:17}) to determine these unknowns and enforce the tangential continuity of the electric field at the interface $S_p$. Substituting (\ref{eq:10}) and (\ref{eq:12}) into (\ref{eq:17}), multiplying the result by $\hat{w}\times\textbf{e}_{t,\alpha}$, and taking the inner product yields
\begin{equation}\label{eq:21}
\begin{split}
[m(&\textbf{e}_{t,\alpha},(\bar{\bar{\textbf{F}}}_{1,p}^{+})^{t})
+m(\textbf{e}_{t,\alpha},(\bar{\bar{\textbf{F}}}_{1,p}^{-})^{t})\tilde{\bar{G}}_{1,2}^{(p)}\bar{\bar{P}}_{1,p}^{+}]\textbf{A}_{1,p}^{\textrm{scat}}\\
&+m(\textbf{e}_{t,\alpha},(\bar{\bar{\textbf{F}}}_{1,p}^{-})^{t})\bar{\bar{P}}_{1,p}^{-}\textbf{B}_{1,p}^{\textrm{inc}}
= m(\textbf{e}_{t,\alpha},\hat{w}\times\textbf{E}_{|_{S_p}})
\end{split}
\end{equation}
The bilinear functions $c(\cdot,\cdot)$, $a(\cdot,\cdot)$, $s_{0}(\cdot,\cdot)$, $s_{p}(\cdot,\cdot)$, $m(\cdot,\cdot)$ and the inner product $(\cdot,\cdot)$ are defined in Appendix 5.1.

\subsection{Discretization and Solution}

As described in \cite{Liu2020feb2}, the 2.5-D MFEM is employed to solve waveguide eigenmodes in $LM_{p}$, eliminating all spurious modes.
The 3-D MFEM with the tree-cotree technique is utilized to discretize the 3-D scattering region $\Omega$, ensuring both tangential continuity and divergence-free conditions of the electric field without additional unknowns, thus providing stable and accurate solutions.

\emph{1) Discretization for waveguide eigenvalue problem:} In the 2.5-D MFEM, the eigenfunctions $\textbf{e}_{t}^{(n)}$ and $\textbf{h}_{t}^{(n)}$ are approximated using the LT/QN edge basis functions $\textbf{N}_{\alpha}$ that enforce the tangential continuity \cite{Liu2020feb2}.
This formulation yields
\begin{equation}\label{eq:22}
\textbf{e}_{t}^{(n)} = \sum_{\alpha=1}^{N_{e}}u_{\alpha}\textbf{N}_{\alpha}, ~~\textbf{h}_{t}^{(n)} =\sum_{\alpha=1}^{N_{e}}h_{\alpha}\textbf{N}_{\alpha}
\end{equation}
where $N_{e}$ denotes the total number of edge degrees of freedom (egde-DoFs) on the waveguide cross-section;
$u_{\alpha}$ and $h_{\alpha}$ are the unknowns of the eigenfunctions $\textbf{e}_{t}^{(n)}$ and $\textbf{h}_{t}^{(n)}$, respectively.
The corresponding longitudinal components $e_{w}^{(n)}$ and $h_{w}^{(n)}$ can also be expanded as
\begin{equation}\label{eq:23}
e_{w}^{(n)} = \sum_{\alpha=1}^{M_{n}}w_{\alpha}\phi_{\alpha}, ~~h_{w}^{(n)} =\sum_{\alpha=1}^{M_{n}}w_{h,\alpha}\phi_{\alpha}
\end{equation}
where $\phi_{\alpha}$ is the second-order Lagrange basis function \cite{Monk2003}; $M_{n}$ is the total number of nodal degrees of
freedom (nodal-DOFs); $w_{\alpha}$ and $w_{h,\alpha}$ are the unknowns of $e_{w}^{(n)}$ and $h_{w}^{(n)}$, respectively.

Substituting (\ref{eq:22}) and (\ref{eq:23}) into the variational formulation corresponding to (\ref{eq:1})-(\ref{eq:3}) yields the associated generalized eigenvalue problem
\begin{equation}\label{eq:24}
\begin{bmatrix}
k_{0}^{2}\bar{\bar{C}}^{'}-\bar{\bar{A}}^{'}& \bar{\bar{B}}_{1}^{'}\\
\bar{\bar{D}}^{'} & \bar{\bar{M}}^{'}
\end{bmatrix}
\begin{bmatrix}
\textbf{u}\\
\textbf{w}
\end{bmatrix}
=(k_{z,h})^{2}
\begin{bmatrix}
\bar{\bar{B}}_{2}^{'}& O\\
O & O
\end{bmatrix}
\begin{bmatrix}
\textbf{u}\\
\textbf{w}
\end{bmatrix}
\end{equation}
where $k_{z,h}$ denotes the approximate eigenvalue; $\textbf{u} =[u_{1}, u_{2}, \ldots ,u_{N_{e}}
]$ and $\textbf{w}=[w_{1}, w_{2}, \ldots, w_{M_{n}}]$ are the unknowns of the approximate fields $\textbf{e}_{t}^{(n)}$ and $e_{w}^{new}$, respectively; the matrices in (\ref{eq:24}) can be found from the Appendix of \cite{Liu2020feb2}. Therefore, once the eigenmodes are obtained from (\ref{eq:24}), the EM fields in each layer within the inhomogeneous layered media $LM_{p}$ can be derived from (\ref{eq:4})-(\ref{eq:13}).

\emph{2) Discretization for 3-D scattering region:} To enforce both tangential continuity and the divergence-free conditions of the electric field, discretizing (14) with edge elements alone is insufficient. Invoking Gauss's law would enlarge the system matrix \cite{Venkatarayalu2006}. We therefore adopt the tree-cotree MFEM, which splits the electric field unknowns into cotree and tree edge components  \cite{Wang2023}. The tree-edge component is approximated by gradients of nodal basis functions on free nodes, representing the null space of $\mathbb{H}(\textrm{curl})$ and enforcing the material-weighted divergence-free conditions.
The cotree-edge component is expressed through the tangential edge basis functions. Consequently, the electric field $\textbf{E}$ can be first expanded as
\begin{equation}\label{eq:25}
\textbf{E} = \sum_{i=1}^{N_{c}}e_{i}^{c}\bm{\Phi}_{i}^{c} + \sum_{j=1}^{N_{n}}e_{j}\nabla\phi_{j}
\end{equation}
where $\bm{\Phi}_{i}^{c}$ and $\phi_{j}$ are the edge and nodal basis functions, respectively; $N_{c}$ and $N_{n}$ are the number of cotree edges and free nodes.
In fact, $\bm{\Phi}_{i}^{c}$ and $\nabla\phi_{j}$ span a finite element space $\textbf{W}_{h}=\textrm{span}\{\bm{\Phi}_{1}^{c},\ldots\bm{\Phi}_{N_c}^{c},\nabla\phi_{(N_c +1)}, \ldots, \nabla\phi_{(N_c+N_n)}\}\subset \mathbb{H}(\textrm{curl})$.
Second, inserting (\ref{eq:22}) and (\ref{eq:25}) into (\ref{eq:20}) and (\ref{eq:21}), we can obtain
\begin{equation}\label{eq:26}
\begin{split}
\bar{\bar{K}}\textbf{e}-c_{0}\sum_{p=1}^{P}\tilde{\bar{C}}_{S_{p}}\textbf{A}_{1,S_{p}}^{\textrm{scat}} = \textbf{s}+c_{0}\sum_{p=1}^{P}\textbf{b}_{S_{p}}
\end{split}
\end{equation}
\begin{equation}\label{eq:27}
\begin{split}
\bar{\bar{Y}}_{S_{p}}\textbf{e}_{S_{p}} - \bar{\bar{U}}_{S_{p}}\textbf{A}_{1,p}^{\textrm{scat}}= \textbf{b}_{S_{p}}^{'}
\end{split}
\end{equation}
where the matrices and source vectors in (\ref{eq:26}) and (\ref{eq:27}) are shown in detail in Appendix 5.2;
the unknowns $\textbf{e}=\{e_{1}^{c},e_{2}^{c},\ldots,e_{N_{c}}^{c},e_{1},e_{2},\ldots,e_{N_{n}}\}^{t}$ and $\textbf{e}_{S_{p}}$ is the subset of $\textbf{e}$ at the interface $S_{p}$.
Finally, the unknowns $\textbf{e}$ and $\textbf{A}_{1,p}^{\textrm{scat}}$ are achieved by solving the systems (\ref{eq:26}) and (\ref{eq:27}), thereby obtaining the electric field in the 3-D scattering region via (\ref{eq:25}).

\emph{3) Determining the fields in $LM_{p}$:} Using $\textbf{A}_{1,p}^{\textrm{scat}}$ as the initial recursion vector, the amplitude $\textbf{A}_{n,p}^{\textrm{scat}}$ of the $n$-th layer in $LM_{p}$ is first analytically obtained from the recursive relation (\ref{eq:7}).
Second, following a procedure similar to (\ref{eq:10}) and (\ref{eq:11}), the scattered fields $\textbf{E}_{n,t;p}^{\textrm{scat}}$ and $\bar{\bar{\mathfrak{R}}}\textbf{H}_{n,t;p}^{\textrm{scat}}$ can be derived from (\ref{eq:4}) and (\ref{eq:5}) for the $n$-th layer.
Finally, the amplitude $\textbf{B}_{n,p}^{\textrm{inc}}$ is acquired by using the initial vector $\textbf{B}_{1,p}^{\textrm{inc}}$ defined by (\ref{eq:12}) and (\ref{eq:13}),
then $\textbf{E}_{n,t;p}^{\textrm{inc}}$ and $\bar{\bar{\mathfrak{R}}}\textbf{H}_{n,t;p}^{\textrm{inc}}$ can be expressed via (\ref{eq:8}) and (\ref{eq:9}).
Once the incident and scattered fields are determined, the total field can be reconstructed for each layer of the layered media $LM_{p}$.

Notably, the systems (\ref{eq:26}) and (\ref{eq:27}) form a $(N_c+N_n+m)\times(N_c+N_n+m)$ matrix, where $(N_c+N_n)$ denotes the total unknowns in the 3-D scattering region and $m$ is the number of waveguide modes utilized in the layered media.
Since $m$ is typically much smaller than $(N_c+N_n)$, it can be neglected. Thus, compared to the full-domain methods, the proposed hybrid method significantly reduces unknowns and computational cost.

\section{Numerical Experiments}

This section presents five numerical examples to validate the accuracy, efficiency, and capability of the proposed hybrid numerical method (HNM).
Results are compared with the commercial 3-D FEM solver COMSOL via port S-parameters, defined as
\begin{equation} \label{eq:28}
|S_{\alpha p}|= 10\log(|P_{\alpha,r}/P_{p,in}|)
\end{equation}
where $P_{\alpha,r}= \int_{S_{\alpha}}\textbf{E}_{L_{\alpha},t;\alpha}^{\textrm{scat}}\times (\textbf{H}_{L_{\alpha},t;\alpha}^{\textrm{scat}})^{}\cdot d\textbf{s}$ and $P_{p,in}=\int_{S_{p}}\textbf{E}_{L_{p},t;p}^{\textrm{inc}}\times (\textbf{H}_{L_{p},t;p}^{\textrm{inc}})^{}\cdot d\textbf{s}$ denote the reflected and incident powers at ports $\alpha$ and $p$ ($\alpha, p=1,2,\ldots P$), respectively. The relative errors across examples range from $0.17\%$ to $1.48\%$.

The number of modes required depends on transverse inhomogeneity, frequency, and port-scattering-region coupling. Strong inhomogeneity or anisotropy demands more modes; short scattering regions require sufficient evanescent modes to capture near-field effects. Table \ref{example0} summarizes relative errors for all examples using different mode counts, confirming that the selected numbers are sufficient for convergence.
\begin{table}[h]
\renewcommand{\arraystretch}{1.1}
\caption{Relative Errors of S-parameters for Different Examples and Modes}
\centering
\begin{tabular}{ccccc}
\hline
Example & Modes & $|S_{11}|$ (\%) & $|S_{21}|$ (\%) & $|S_{31}|$ (\%) \\
\hline
A & 50 / 100 & 1.22 / 0.77 & 1.82 / 1.48 & - \\
B & 4 / 8 & 0.97 / 0.97 & 0.17 / 0.17 & - \\
C & 50 / 80 & 0.57 / 0.46 & 0.82 / 0.51 & 0.84 / 0.52 \\
D & 10 / 20 & 0.20 / 0.19 & 0.32 / 0.31 & 0.46 / 0.44 \\
E & 50 / 80 & 1.49 / 1.43 & 0.91 / 0.90 & 0.92 / 0.91 \\
\hline
\end{tabular}
\label{example0}
\end{table}

For implementation, COMSOL generates a 3-D mesh of the entire model; the scattering region $\Omega$ mesh is exported to 3-D MFEM, and the 2-D interface meshes at $S_p$ to 2.5-D MFEM for solving eigenmodes. Compared with COMSOL, HNM shows superior CPU time, memory usage, and degrees of freedom (DOF), with consistent accuracy. The CPU time is averaged over all frequency points, including eigenmode calculation. Linear systems in(\ref{eq:26}) and (\ref{eq:27}) are solved with the direct solver PARDISO \cite{Schenk2004} (also used in COMSOL), on a desktop with an Intel Xeon Gold 6342 CPU $@$ 2.80 GHz and 1.0 TB RAM.

\subsection{Multi-Layered Media Nested With Scatterer}

\begin{figure}[h]
\centering
{
\includegraphics[width=0.7\columnwidth,draft=false]{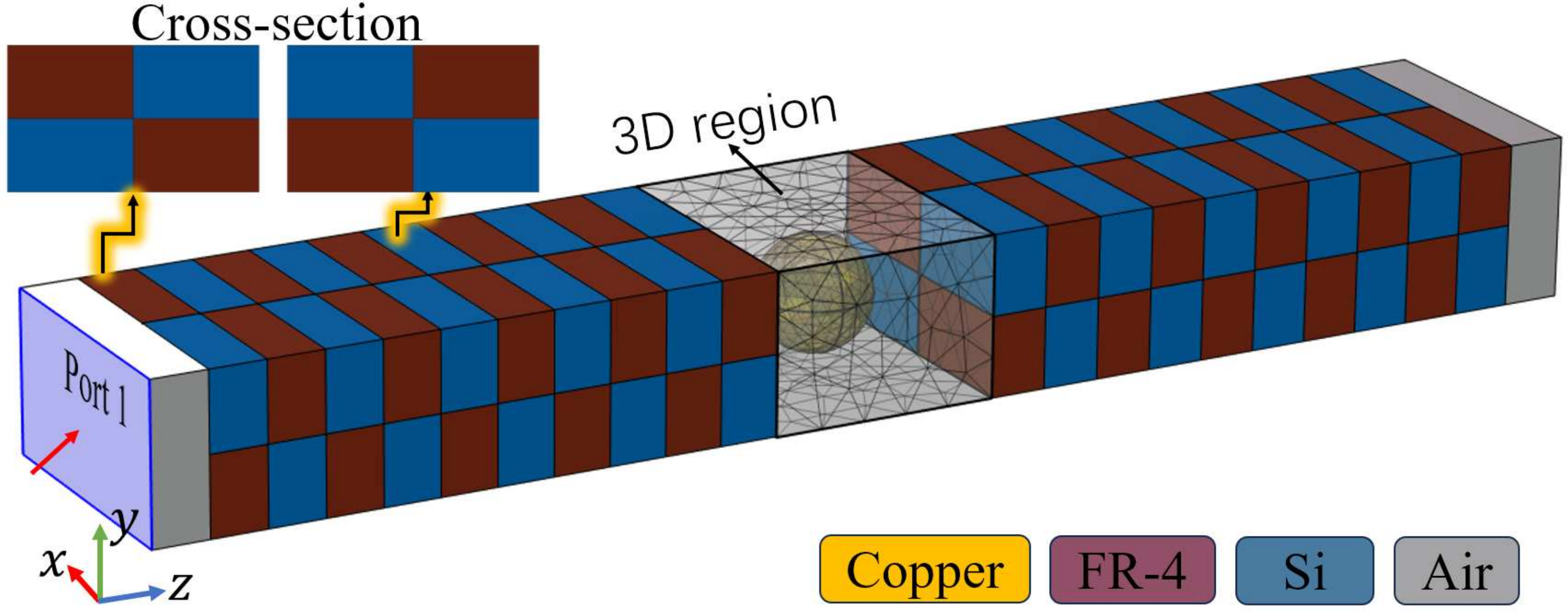}
}
\caption{Schematic of multi-layered inhomogeneous media with a 3-D scattering region containing a metallic sphere.
The top, bottom, front, and rear boundaries are Bloch periodic.
A plane wave at $30^{\circ}$ to the $z$-axis illuminates the structure from Port 1.}
\label{sketch2}
\end{figure}

We first consider a multi-layered inhomogeneous structure with a copper spherical scatterer as shown in Fig. \ref{sketch2}, which cannot be adequately handled by the traditional hybrid NM/FEM using orthogonal TE/TM modes.
The structure comprises alternating checkerboard layers (silicon and FR-4 blocks) and two air layers serving as input/output ports, mimicking a packaged antenna or RF module in a multilayered PCB environment.
Every layer is $24\,\textrm{cm}\times 14\,\textrm{cm}\times 10\,\textrm{cm}$.
The scattering region includes an air layer of $24\,\textrm{cm}\times 14\,\textrm{cm}\times 20\,\textrm{cm}$ and a copper sphere of radius $5\,\textrm{cm}$.
Bloch periodic boundaries are applied to the outer surfaces.
A plane wave at $30^{\circ}$ to the $z$-axis excites the structure from Port 1 over 0.6 to 1 GHz.
The relative permittivity are $\epsilon_r = 11.7$ (silicon) and $\epsilon_r = 4.2$ (FR-4 ); copper conductivity is $5.8\times10^7\,\textrm{S/m}$.

\begin{figure}[h]
  \centering
  {
   \includegraphics[width=0.7\columnwidth,draft=false]{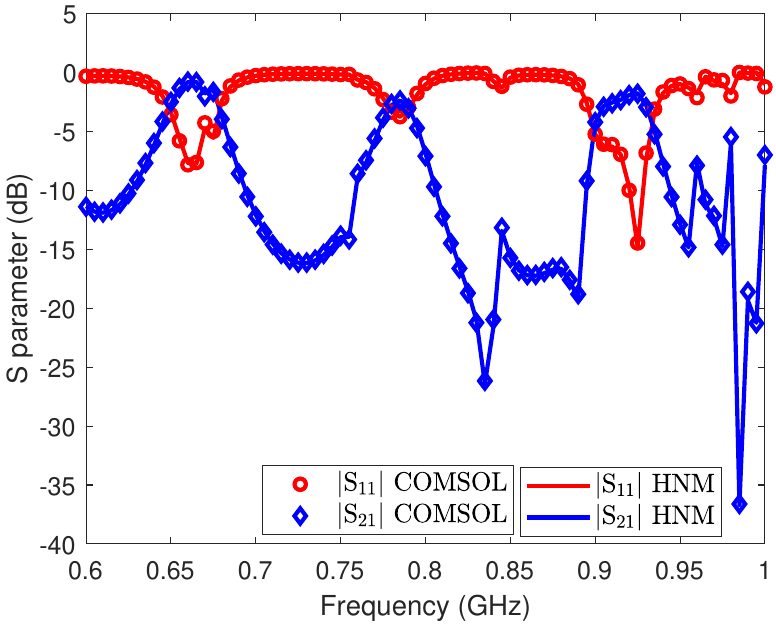}
   }
 \caption{S-parameters at the ports of multi-layered inhomogeneous media as shown in Fig. \ref{sketch2}, where sampling is performed at the same 81 frequency points for both COMSOL and HNM.}
\label{Fig4}
\end{figure}

The scattering region is discretized with a 3-D mesh for the MFEM, while the layered-media EM fields are described via the NMM with second-order basis functions in both HNM and COMSOL. 100 modes are used for the NMM due to the interleaved inhomogeneous media and Bloch periodic boundaries in each layer.
The S-parameters at Port 1 and Port 2 are shown in Fig. \ref{Fig4}; computation cost comparisons are summarized in Table \ref{example1}, where the HNM is 2.09 times faster, uses 6.64 times fewer DOFs, and requires 5.54 times less memory than COMSOL.
The S-parameters obtained by the HNM and COMSOL match well,
with the relative errors summarized in Table \ref{example0}.
\begin{table}[h]
\renewcommand{\arraystretch}{1.1}
\caption{Computational Costs for Multi-Layered Inhomogeneous Media}
\centering
\begin{tabular}{ccccc}
\hline
Solver   & DOF     & Time (s) & Memory (GB)\\
\hline
HNM     & 109566    & 15.72 & 3.89\\

COMSOL   & 727358    & 32.93  & 21.54\\
\hline
\end{tabular}
\label{example1}
\end{table}

\subsection{Bend Waveguide Model}

\begin{figure}[h]
  \centering
  {
   \includegraphics[width=0.7\columnwidth,draft=false]{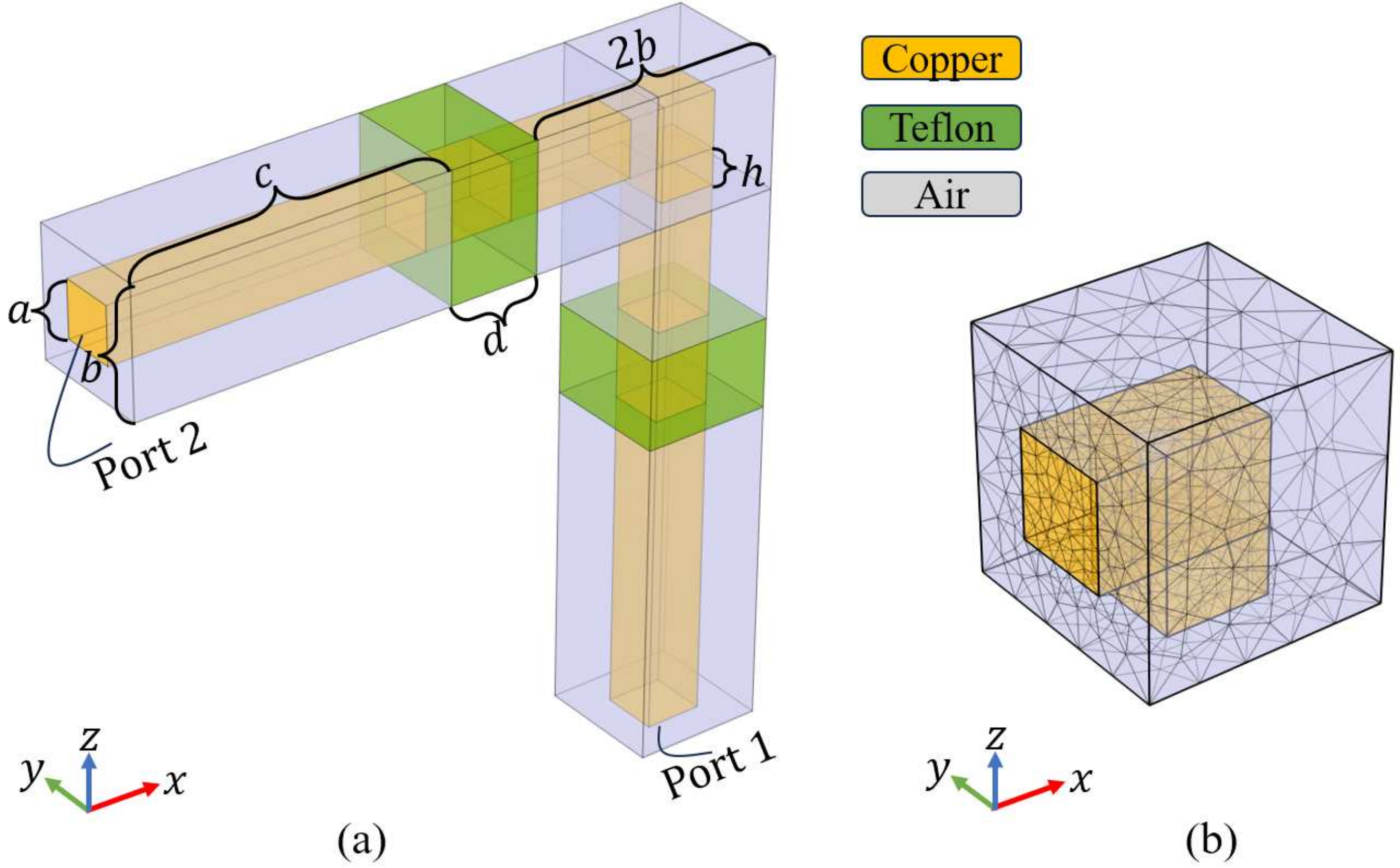}
   }
 \caption{Geometry of bend waveguide model. (a) The model consists of a $90^{\circ}$ curved copper conductor and two Teflon hollow cubic supports. Air-filled regions surround the copper conductor except for the Teflon supports. All outer boundaries are PEC except at the two ports. (b) Mesh of the $90^{\circ}$-bend section for the MFEM.}
\label{sketch3}
\end{figure}

As microwave circuit integration advances, bent waveguides have become essential for miniaturization and component interconnection in compact designs,
 and are widely used in mode converters, filters, couplers, and antenna systems.
Accurate characterization of wave propagation in a bent waveguide is therefore important.
To verify that our method supports TEM excitation, we apply the HNM to compute S-parameters for the $90^{\circ}$ bent waveguide model as shown in Fig. \ref{sketch3}, a verification not achievable  by the traditional hybrid NM/FEM based on the scalar Helmholtz equation.

The model (Fig. 5(a)) consists of a $90^{\circ}$ curved copper conductor and two Teflon hollow cubic supports ($\epsilon_{r}=2.08$, $\tan\delta = 0.001$).
The inner conductor has a square cross-section $a\times a$. Each support is a hollow square column with outer side $b$, inner side $a$, and thickness $d$, connected to two air enclosures of dimensions $b\times b\times c$ and $b\times b\times b$, respectively.
In the bend section, the copper conductor has a long side $(a+h)$ and a short side $h$, enclosed by a cubic air box of edge length $b$.
The geometric parameters are: $a=4.4$ mm, $b=10$ mm, $c=25$ mm, $d=7$ mm, $h=2.8$ mm.
Outer boundaries are PEC except at the two ports.

\begin{figure}[h]
  \centering
  {
   \includegraphics[width=0.7\columnwidth,draft=false]{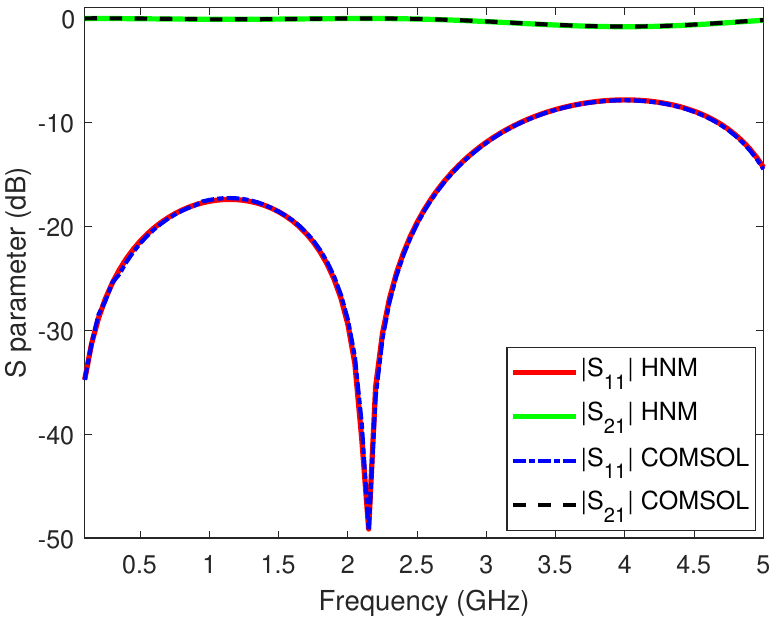}
   }
 \caption{S-parameters at the ports of bend waveguide model. The same frequency sampling (99 frequency points) is adopted for both HNM and COMSOL.}
\label{Fig6}
\end{figure}

The HNM simulates the bend section with a 3-D mesh as shown in Fig. 5(b), while the supports, conductors, and air enclosures are modeled as via the NMM. A TEM mode is supported at both ports; port 1 provides the excitation over $0.1\sim 5$ GHz.
Since the TEM mode is propagating and all others are evanescent, 4 modes are used in the NMM.
As shown in Table \ref{example2}, the HNM is 5.46 times faster than COMSOL,
while COMSOL requires approximately 12.87 times more DOFs and 15.78 times more memory.
Fig. \ref{Fig6} shows good agreement between the two solvers for S-parameters $|S_{11}|$ and $|S_{21}|$, with the relative errors listed in Table \ref{example0}.
\begin{table}[h]
\renewcommand{\arraystretch}{1.1}
\caption{Computational Costs for Bend Waveguide}
\centering
\begin{tabular}{ccccc}
\hline
Solver   & DOF     & Time (s) & Memory (GB)\\
\hline
HNM     & 93380    & 4.91 & 2.10\\

COMSOL   & 1201378 & 26.80  & 33.13\\
\hline
\end{tabular}
\label{example2}
\end{table}

\subsection{Multiple Scatterers Model}

\begin{figure}[!t]
  \centering
  \subfigure[]{
    \label{sketch4a}
   \includegraphics[width=0.5\columnwidth,draft=false]{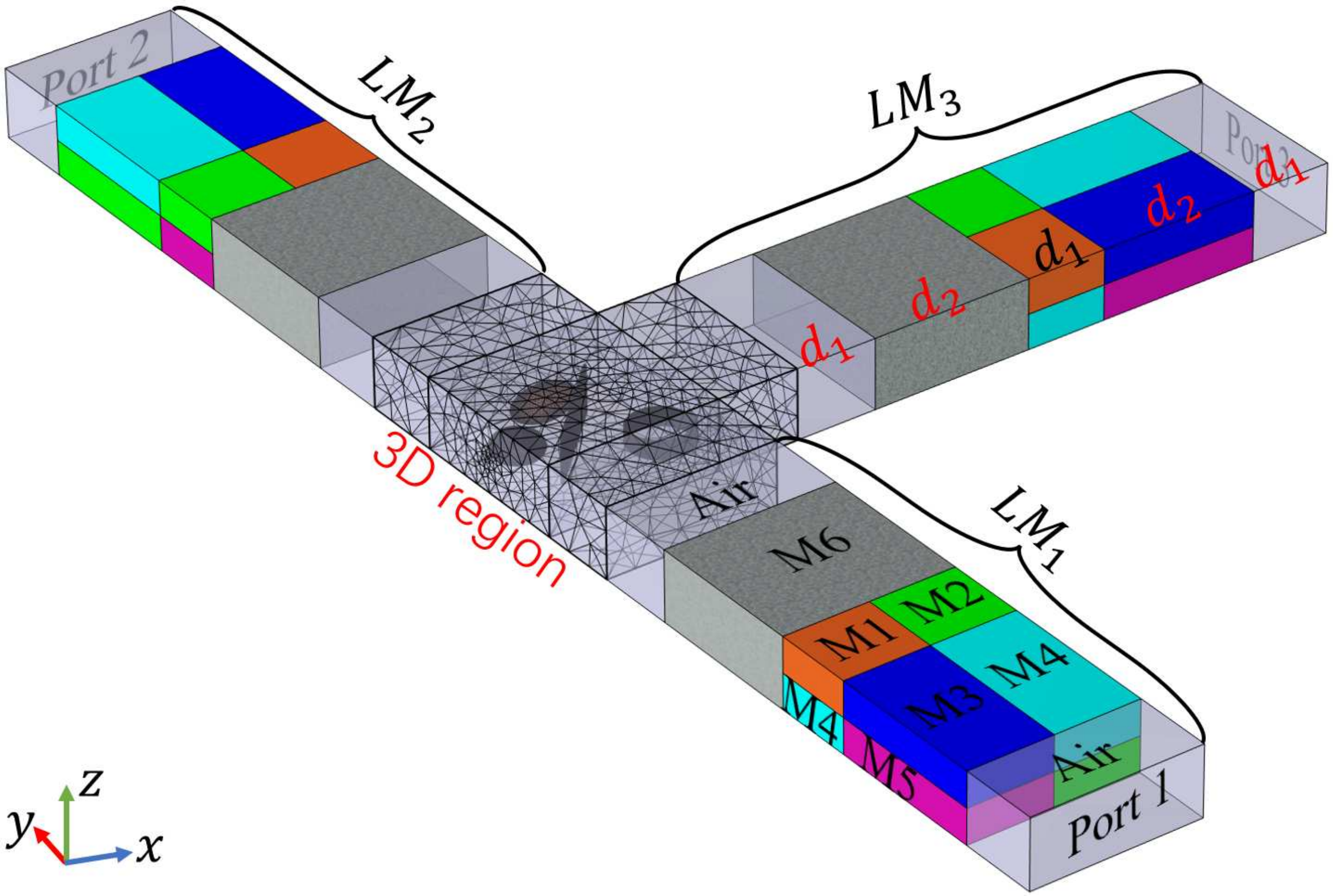}}
     \subfigure[]{
    \label{sketch4b}
   \includegraphics[width=0.4\columnwidth,draft=false]{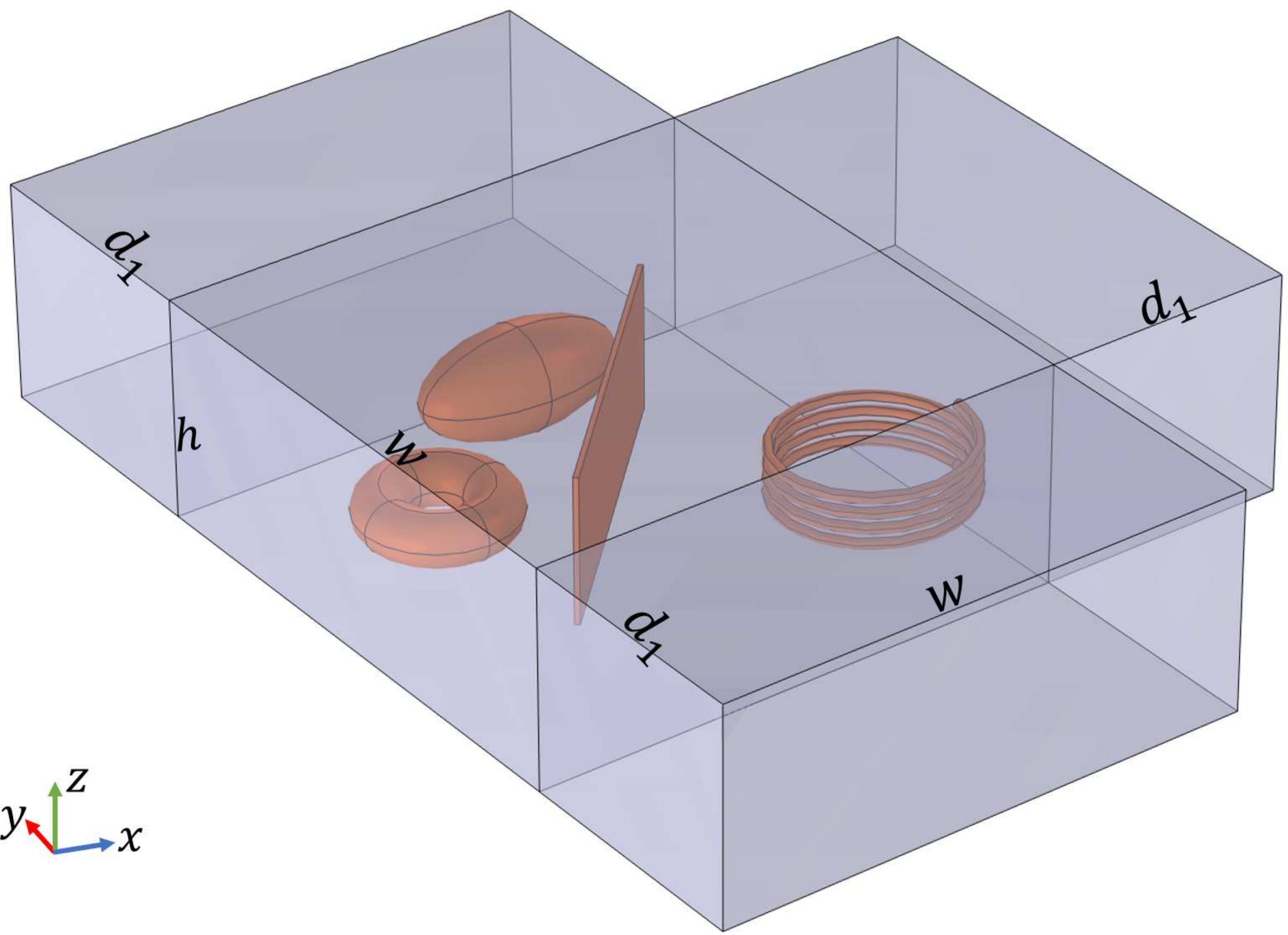}}
   \caption{Geometry of the multiple scatterers model. (a) The model consists of three layered media $LM_{i}$ ($i=1,2,3$) and a scattering region. $LM_2$ and $LM_3$ are obtained by rotating $LM_1$ by $90^{\circ}$ and $180^{\circ}$ around the $z-$axis, respectively. Their constituent blocks are filled with material $M_{\alpha}$ and air ($\alpha=1,2,\ldots,6$).
   (b) Detailed view of the scattering region. A ridge air cavity contains metallic components: a cube, a solenoid, a torus, and an ellipsoid. All outer boundaries are PEC except at the three ports. Port 1 is excited by the $\textrm{TE}_{10}$ waveguide mode.}
\end{figure}
We now validate the HNM for multiple scatterers using the model in Fig. \ref{sketch4a}, which consists of three inhomogeneous layered media $LM_{i}$ ($i=1,2,3$) coupled with a 3-D ridge-scattering air domain containing four metallic scatterers: a cube, helix, ellipsoid, and torus.
This configuration can simulate a 3-D antenna array unit cell with parasitic elements fed by layered inhomogeneous waveguides. Such problems are difficult for conventional hybrid NM/FEM, which rely on orthogonal TE/TM modes or empty cascaded waveguides.

As shown in Fig. \ref{sketch4b}, the scatterer geometries are: cube at $(w/2,dt+w/2,h/2)$, dimensions $d_0\times 2w/3\times 2h/3 $; four-turn helix at $(4w/5,dt+w/5,h/3)$, major radius 1.8 mm, minor radius 0.1 mm, pitch 0.3 mm; ellipsoid at $(w/2,dt+3w/4,h/2)$, semi-axes 2, 1, 1 mm; torus at $(w/6,dt+w/2,h/2)$, major radius 1 mm, minor radius 0.5 mm.
The parameters: $d_0=0.1$ mm, $d_1=5$ mm, $d_2=10$ mm, $dt=40$ mm, $w=10.67$ mm, $h=4.318$ mm. The relative permittivities are: $\epsilon_r =5.2$ (M1), $\epsilon_r =7$ (M2), $\epsilon_r =9(1-0.005j)$ (M3), $\epsilon_r = 4$ (M4), $\epsilon_r = 5$ (M5), $\bar{\bar{\epsilon}}_r = diag(4,3,1)$ (M6).

\begin{figure}[h]
  \centering
  {
   \includegraphics[width=0.7\columnwidth,draft=false]{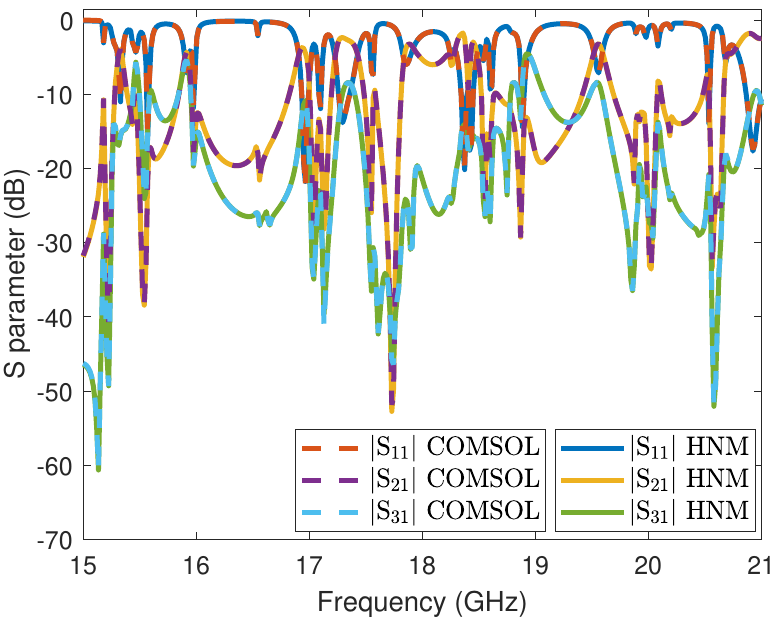}
   }
 \caption{S-parameters at the ports of the multiple-scatterers model in Fig. 7, where sampling is performed at the same 401 frequency points for both COMSOL and HNM.}
\label{Fig8}
\end{figure}

Simulations use second-order basis functions in both HNM and COMSOL.
The layered-media fields are constructed with 80 modes in the NMM due to possible mode conversion from inhomogeneity. A $\textrm{TE}_{10}$ mode excitation at Port 1 sweeps $15\sim21\,\textrm{GHz}$ over 201 Tables \ref{example3} shows the HNM is 2.00 times faster, uses 3.16 times fewer DOFs, and 2.66 times less memory than COMSOL.
The S-parameters from both solvers agree closely, as shown in Fig. \ref{Fig8}; the relative errors are listed in Table \ref{example0}.
Having covered plane wave, TEM, and $\textrm{TE}_{10}$ excitations in inhomogeneous lossy/anisotropic layered media with single or multiple scatterers, we further demonstrate the capability of HNM for dispersive media and arbitrarily-shaped ports.

\begin{table}[h]
\renewcommand{\arraystretch}{1.1}
\caption{Computational Costs for Multiple Scatterers Model}
\centering
\begin{tabular}{ccccc}
\hline
Solver   & DOF     & Time (s) & Memory (GB)\\
\hline
HNM     & 619452    &40.20   &16.32\\

COMSOL   & 1958732  & 80.40 & 43.42\\
\hline
\end{tabular}
\label{example3}
\end{table}

\subsection{Millimeter-Wave Circulator Model}

\begin{figure}[h]
  \centering
  \subfigure[]{
    \label{sketch5a}
   \includegraphics[width=0.5\columnwidth,draft=false]{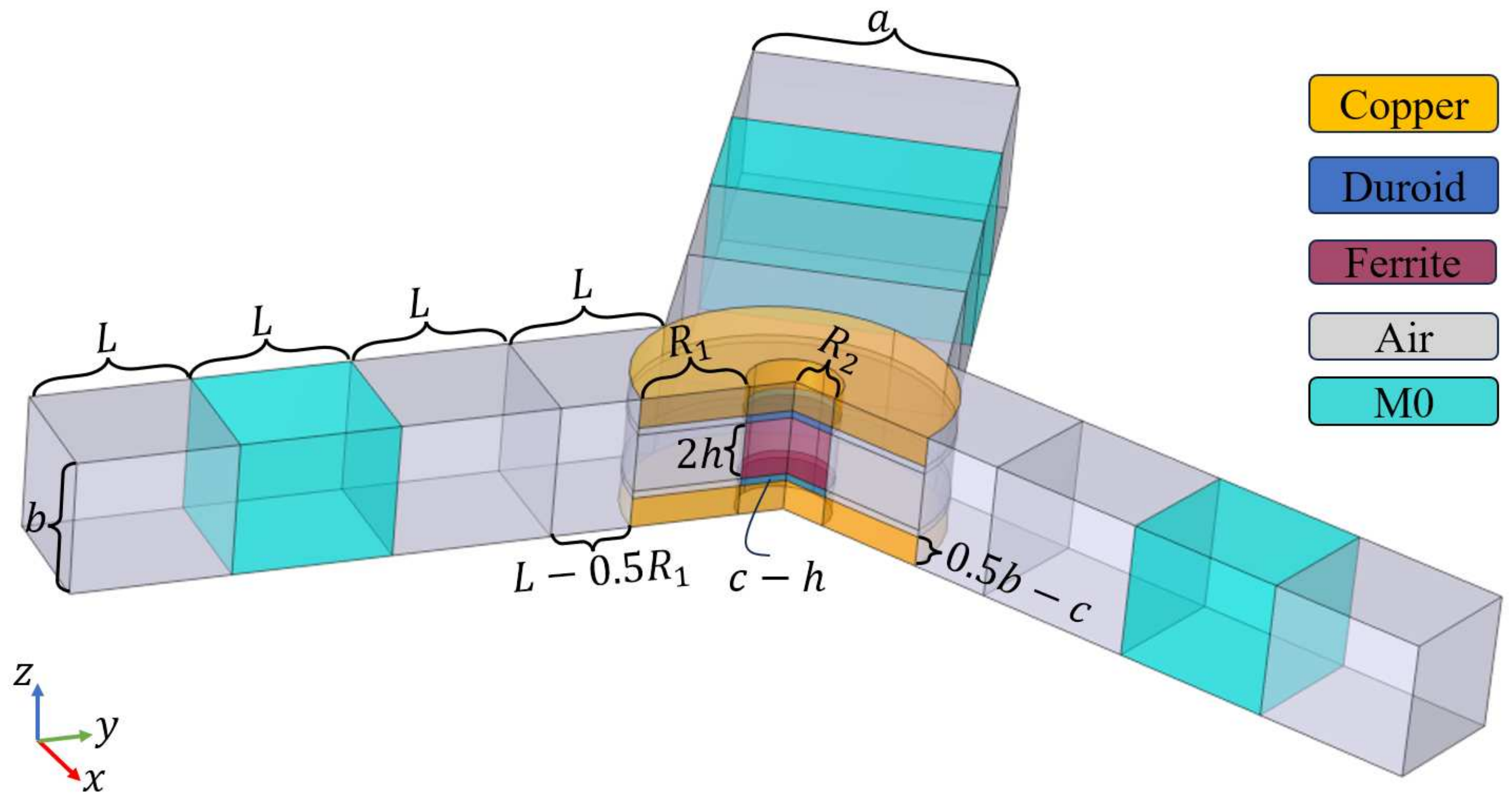}}
     \subfigure[]{
    \label{sketch5b}
   \includegraphics[width=0.4\columnwidth,draft=false]{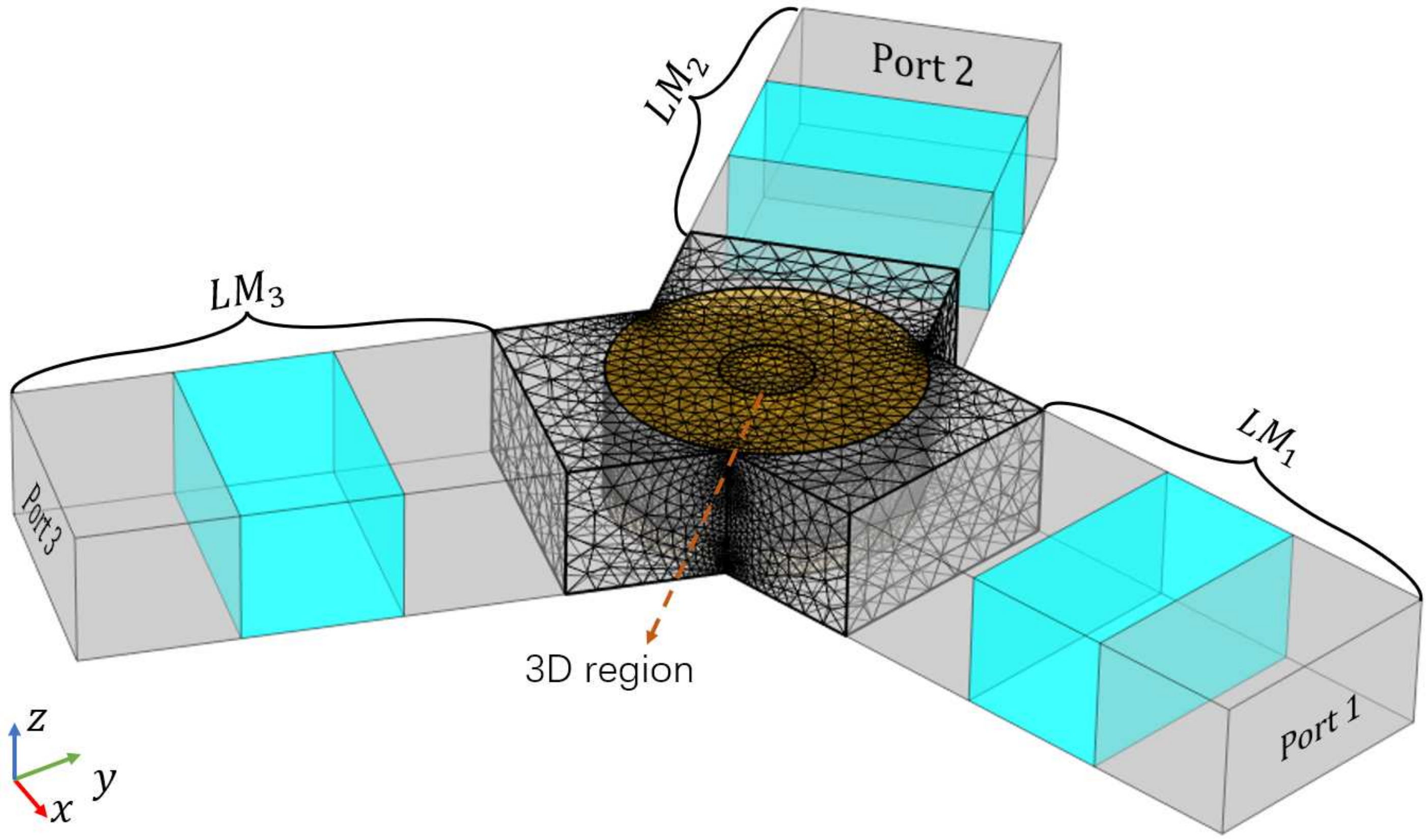}}
   \caption{Geometry of the millimeter-wave circulator model. (a) Sectional view. (b) The circulator model consists of three layered media $LM_{i}$ ($i=1,2,3$) and a scattering region, with three ports at the extremities of $LM_{i}$.
   All other outer boundaries are PEC.
   The scattering region is meshed with tetrahedron for the MFEM, while $LM_{i}$ is described by the NMM. Port 1 is excited by the $\textrm{TE}_{10}$ waveguide mode.}
\end{figure}

The circulator is a common component in microwave and millimeter systems for directional wave propagation, with applications in antenna duplexers, isolators, and radar systems.
We consider a millimeter-wave circulator model similar to that in Figure 9 of \cite{Liu2002}, sketched in Fig. \ref{sketch5a}.
The model consists of a cylinder and three waveguides.
From top to bottom, the cylinder has five layers: copper, Duroid, strontium ferrite, Duroid, and copper, with heights $(0.5b-c)$, $(c-h)$, $2h$, $(c-h)$, and $(0.5b-c)$, respectively.
The copper radius is $R_{1}=a/\sqrt{3}$; the Duroid and ferrite radii are $R_{2}=1.57h$, with $R_{2} < R_{1}$, surrounded by air.
Parameters: $a=7.11$ mm, $b=3.56$ mm, $c=0.978$ mm, $h=0.78$ mm.

As shown in Fig. \ref{sketch5b}, to facilitate the HNM simulation, the three waveguides are truncated at a distance $L+R_{1}/2$ from the cylinder axis into three saddle-shaped curved hexahedra and three straight hexahedra ($L=4.16$ mm).
These curved hexahedra and the cylinder form a scattering region, meshed with tetrahedron for HNM. The layered media $LM_{i}$ ($i=1,2,3$) of dimensions $a\times3L\times b$ connect to the ports. 10 eigenmodes are used for the NMM, as each layer is homogeneous and field behavior is simple.

\begin{figure}[h]
  \centering
  \subfigure[]{
    \label{Fig10(a)}
   \includegraphics[width=0.6\columnwidth,draft=false]{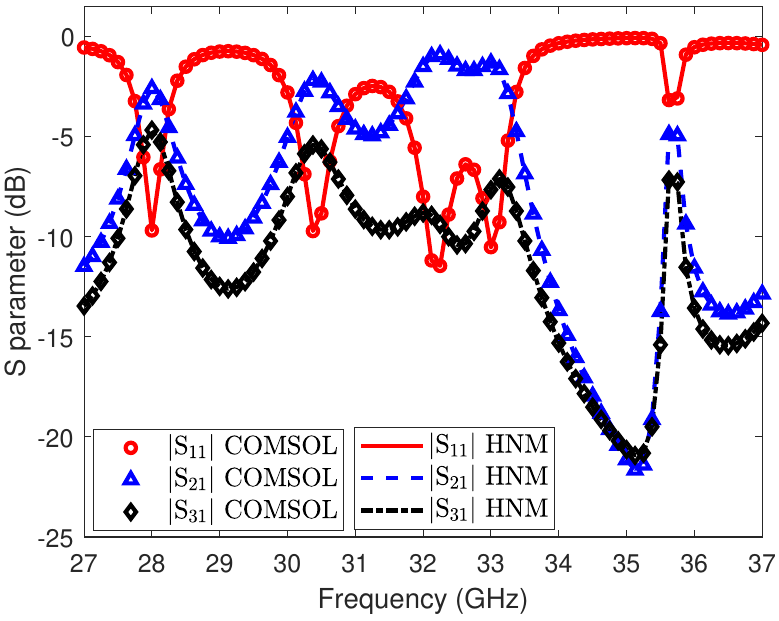}}
   \subfigure[]{
    \label{Fig10(d)}
   \includegraphics[width=0.6\columnwidth,draft=false]{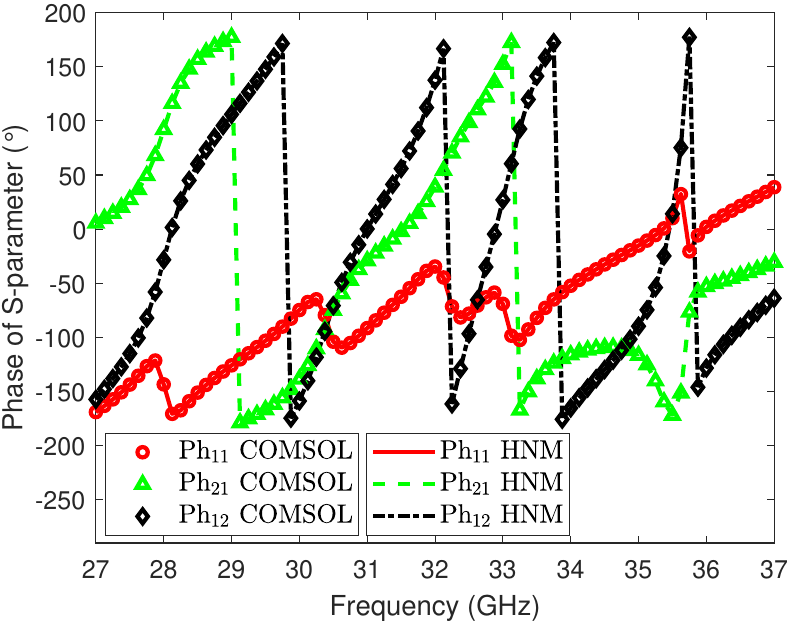}}
   \caption{S-parameters and their phases for the circulator model in Fig. 9. (a) Magnitude of S-parameters. (b) Phase of S-parameters, defined as $\textrm{Ph}=\arctan(Re(S)/Im(S))$. Results from both HNM and COMSOL are shown at 81 frequency points.}
\end{figure}

The relative permittivity of the M0, Duroid and the strontium ferrite are $\epsilon_{M0}= 0.5$, $\epsilon_{D,r} = 2.2$ and $\epsilon_{f,r} = 19.0$, respectively.
Duroid and M0 are non-magnetic; strontium ferrite has permeability
\begin{equation*}
\bar{\bar{\mu}}=\begin{bmatrix} \mu & -j\kappa & 0\\-j\kappa & \mu &0\\0& 0& \mu_{0}\end{bmatrix}\\
\end{equation*}
where $\mu = \mu_{0}+\kappa$, $\kappa = \mu_{0}\frac{\omega_{0}\omega_{m}}{\omega_{0}^2 - \omega^2}$, $\omega_{0}=\mu_{0}\gamma_{e}(H_{A}-M_{r})$, $\omega_{m}=\mu_{0}\gamma_{e} M_{r}$; $\gamma_{e}$ is the electron gyromagnetic ratio, $H_{A} = 50/\pi$ kA/m, and $M_{r} = 1250/\pi$ kA/m.
Simulations run from 27 to 37 GHz using second-order basis functions in HNM and COMSOL.
Tables \ref{example4} shows the HNM is 2.34 times faster, uses 2.21 times fewer DOFs, and 1.64 times less memory than COMSOL.
The S-parameters obtained by the two solvers are compared in Fig. \ref{Fig10(a)}, showing close agreement; the relative errors are listed in Table \ref{example0}.

\begin{table}[h]
\renewcommand{\arraystretch}{1.1}
\caption{Computational Costs for Millimeter-Wave Circulator}
\centering
\begin{tabular}{ccccc}
\hline
Solver   & DOF     & Time (s) & Memory (GB)\\
\hline
HNM     & 1779040    & 137.26   &80.76\\

COMSOL   & 3934911  & 321.22 & 132.44\\
\hline
\end{tabular}
\label{example4}
\end{table}

The model is symmetric under $120^{\circ}$ rotation about the $z$-axis, yielding nonreciprocal behavior; the S-parameters matrix satisfies the corresponding relation at each frequency.
\begin{equation*}
\begin{split}
\begin{bmatrix} |S_{11}| & |S_{12}| & |S_{13}|\\|S_{21}| & |S_{22}| &|S_{23}|\\|S_{31}|& |S_{32}|& |S_{33}|\end{bmatrix}
=\begin{bmatrix} |S_{11}| & |S_{12}| & |S_{21}|\\|S_{21}| & |S_{11}| &|S_{12}|\\|S_{12}|& |S_{21}|& |S_{11}|\end{bmatrix}\\
\end{split}
\end{equation*}
As seen in Fig. \ref{Fig10(d)}, the phases $\textrm{Ph}_{21}$ and $\textrm{Ph}_{12}$ differ significantly, with a relative error of 134.71\%. Nevertheless, HNM and COMSOL agree well, with relative errors of $0.06\%$, $0.10\%$ and $0.17\%$ for $\textrm{Ph}_{11}$, $\textrm{Ph}_{21}$ and $\textrm{Ph}_{12}$, respectively.

\subsection{Simplified Three-Port Divider Model}

\begin{figure}[!t]
  \centering
  \subfigure[]{
    \label{sketch6a}
   \includegraphics[width=0.6\columnwidth,draft=false]{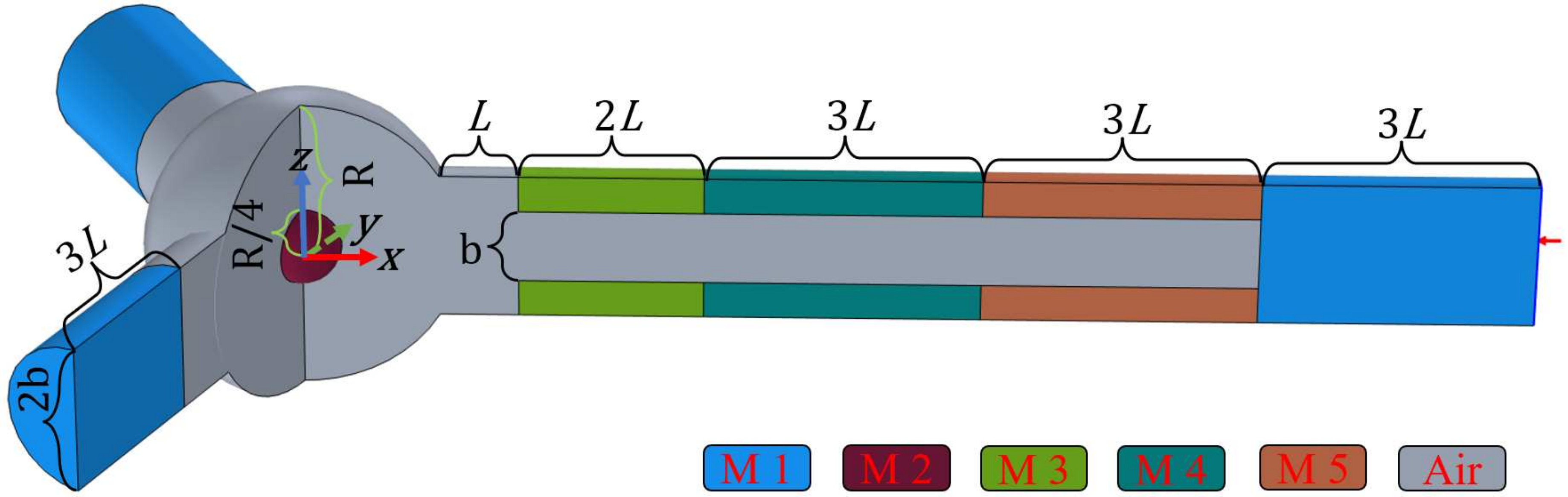}}
     \subfigure[]{
    \label{sketch6b}
   \includegraphics[width=0.5\columnwidth,draft=false]{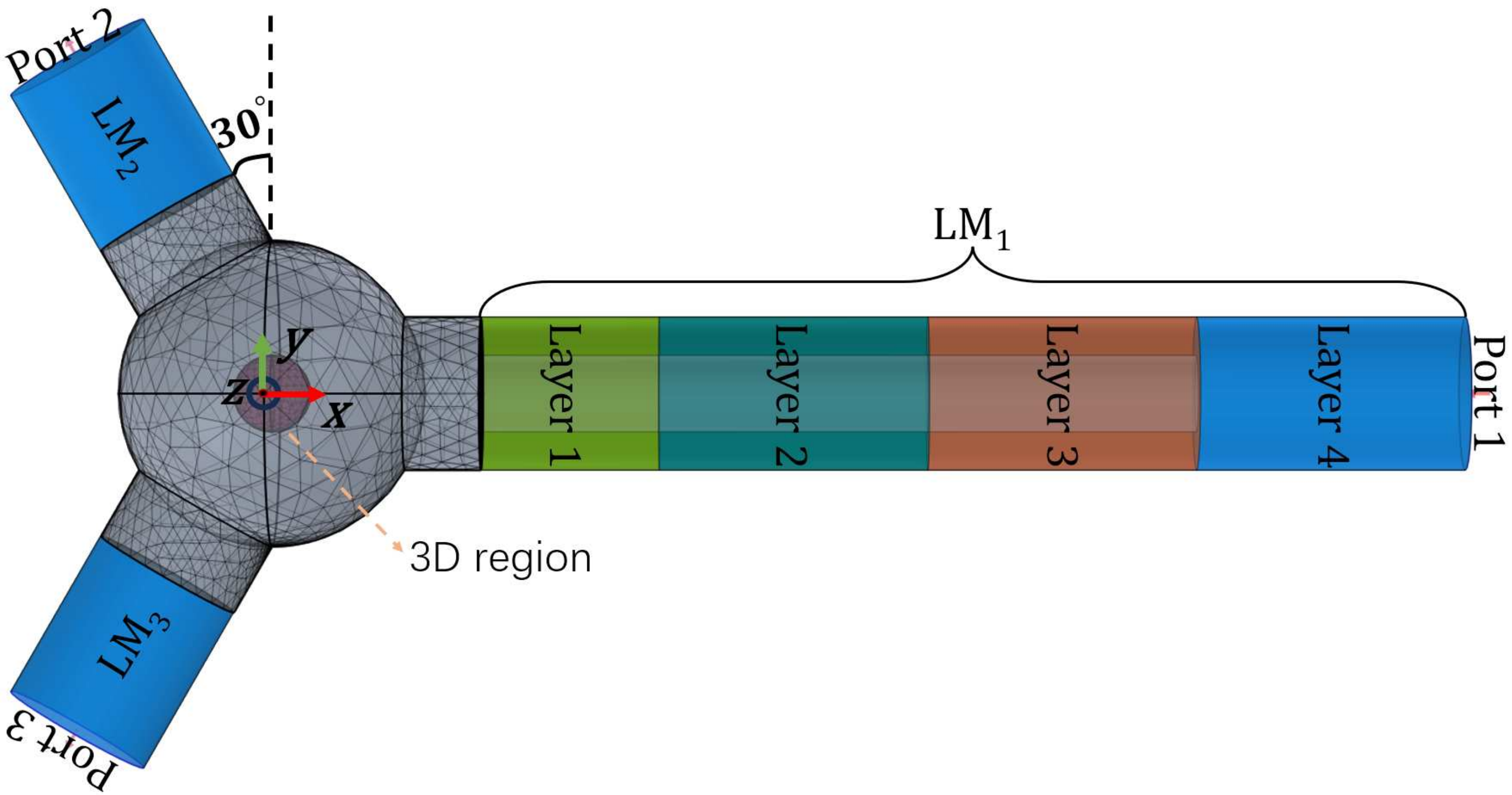}}
   \caption{Geometry of the simplified divider model. (a) Sectional view. (b) The simplified divider model comprises a scattering region and three layered media $LM_{i}$ ($i=1,2,3$).
   The extremities of $LM_{i}$ are set as the ports and the outer boundaries are covered with the PEC.
   The first layered media $LM_{1}$ consists of 4 layers, where layers 1-3 are filled with inhomogeneous media and the fourth layer is filled with a homogeneous medium.
   $LM_{2}$ and $LM_{3}$ each possess a homogeneous layer.}
\end{figure}

The divider is a key component in antenna feed networks, phased arrays, and multi-layer RF systems.
To demonstrate the capability of HNM for multiple inhomogeneous, anisotropic, and lossy layered media with arbitrary cross-sections,
we simulate a simplified three-port divider model as the final example, with its cross-section shown in Fig. \ref{sketch6a}.

The model comprises a scattering region and three waveguide structures.
The scattering region contains a spherical scatterer (radius $R/4$), a spherical shell (radius $R$), and three cylindrical probes connected to the waveguides.
As shown in Fig. \ref{sketch6b}, the first waveguide ($LM_1$) has four layers: the first three are inhomogeneous cylinders with outer radius $b$, inner radius $0.5b$, and heights $2L$, $3L$ and $3L$, respectively; the fourth is a homogeneous cylinder of radius $b$ and height $3L$.
The second and third waveguides ($LM_2$, $LM_3$) are homogeneous cylinders with the same parameters as layer 4.
Parameters: $R=7.11$ mm, $b=3.56$ mm and $L=4.16$ mm.
The scattering region is tetrahedrally meshed for the MFEM; layered media $LM_{i}$ are described by the NMM.

The material distribution is shown in Fig. \ref{sketch6a}: the scatterer is M2, enclosed by an air shell.
In $LM_1$, the first three outer layers are are M3, M4, and M5, respectively, with air cores, forming radially inhomogeneous structures;
the fourth layer is homogeneous M1. $LM_2$ and $LM_3$ are also M1.
All $M_i$ ($i=1,2,\ldots,5$) are non-magnetic, with the relative permittivities of $\epsilon_r =2.2$ (M1), $\epsilon_r =19.0$ (M2), $\epsilon_r =9.8$ (M3), $\epsilon_r = 3.66(1-0.0037j)$ (M4), and $\bar{\bar{\epsilon}}_r = diag(3.0,4.8,3.8)$ (M5).

\begin{table}[h]
\renewcommand{\arraystretch}{1.1}
\caption{Computational Costs for Simplified Circulator Model}
\centering
\begin{tabular}{ccccc}
\hline
Solver   & DOF     & Time (s) & Memory (GB)\\
\hline
HNM     & 322200    & 20.20   &8.16\\

COMSOL   & 674647  & 42.51 & 15.90\\
\hline
\end{tabular}
\label{example5}
\end{table}

\begin{figure}[h]
  \centering
  \subfigure[]{
    \label{Fig12(a)}
   \includegraphics[width=0.6\columnwidth,draft=false]{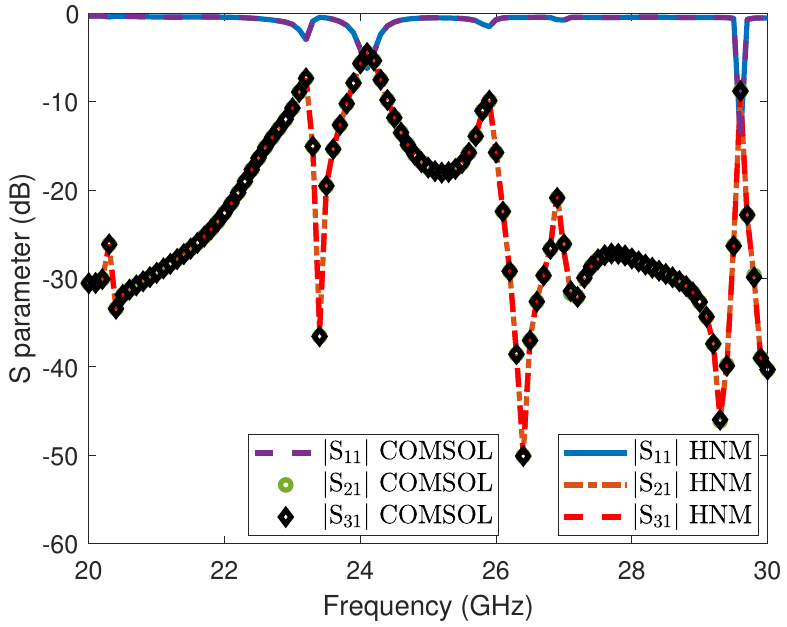}}
   \subfigure[]{
    \label{Fig12(b)}
   \includegraphics[width=0.6\columnwidth,draft=false]{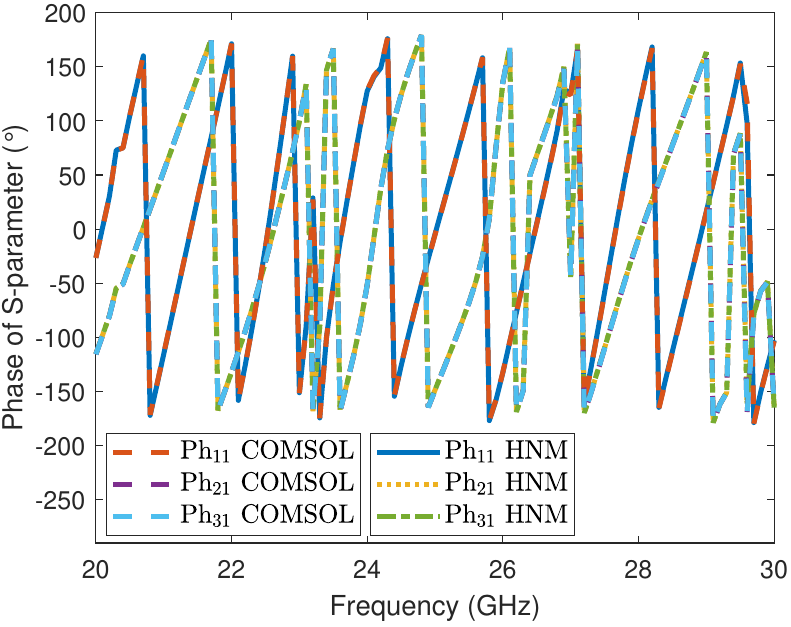}}
   \caption{S-parameters and their phases at the ports of the divider model in Fig. 11. (a) Magnitude of S-parameters. (b) Phase of S-parameters $\textrm{Ph}=\arctan(Re(S)/Im(S))$. The results of both HNM and COMSOL are shown at 101 frequency points.}
\end{figure}

The fundamental $\textrm{TE}_{11}$ mode excites Port 1 over $20\sim30\,\textrm{GHz}$.
The HNM and COMSOL with the second-order basis functions are employed to obtain the S-parameters at the three ports shown in Fig. \ref{Fig12(a)},
where there are 50 modes used by the HNM.
The computational costs are compared in Table \ref{example5}, the HNM is 2.10 times faster, uses 1.95 times less memory, and 2.09 times DOFs than COMSOL.
The S-parameters from both solvers align well, as shown in Fig. \ref{Fig12(a)}, the relative errors are listed in Table \ref{example0}. As illustrated in Fig. \ref{Fig12(b)}, the S-parameter phases obtained by the two methods are also in good agreement, with relative errors of $1.89\%$, $0.91\%$, and $0.92\%$, respectively.
According to Figs. \ref{Fig12(a)} and (b), the symmetry of the simplified divider model along the $x$-axis leads to identical characteristics for the pairs ($|S_{21}|$, $|S_{31}|$) and their corresponding phases, respectively. Since $LM_1$ is filled with inhomogeneous, anisotropic, and lossy layered media, this divider model exhibits unconventional behavior, i.e.,
$|S_{11}|$ is greater than $|S_{21}|$ or $|S_{31}|$.

\section{Conclusion}

A 3-D hybrid method (HNM) is developed for simulating EM fields in structures with a non-layered 3-D region connected to multiple inhomogeneous layered media. It integrates the 3-D NMM with a tree-cotree-based MFEM in one variational framework. The HNM inherits the dimensionality-reduction advantage of both conventional MM/FEM and the pure NMM, while addressing certain aspects of conventional MM/FEM where improvements can be made, including spurious modes, reliance on orthogonal TE/TM modes, and the need for costly matrix inversion in GSM construction. To this end, the HNM adopts a spurious-free 2.5-D MFEM for mode computation, recursively derives reflection matrices without requiring mode orthogonality, and enforces direct tangential field matching as boundary conditions. In the 3-D region, the tree-cotree MFEM enforces tangential continuity and the divergence-free condition without extra unknowns, which also serves as the physical mechanism to maintain accuracy and stability.

Numerical experiments on five models show speedups of 2.00-5.46, DOF reductions of 1.95-12.87, and memory savings of 1.64-15.78 times versus COMSOL, with S-parameter errors of 0.17\%-1.48\%. In fact, the advantage of the HNM becomes more pronounced when the layered media are electrically thick or have complex cross-sections. The method works for plane wave, TEM, $\textrm{TE}_{10}$, and $\textrm{TE}_{11}$ excitations, and for dispersive, anisotropic, lossy, and nonreciprocal media. The examples are chosen for algorithmic validation, but they share essential electromagnetic traits with practical applications such as antenna feeds, RF packages, and PCB transitions. The HNM can also handle multiple scattering regions and layered media. Future work will apply it to antenna feeds, RF packages, and EMC testing.

\section {Appendices}

\subsection{Bilinear Functions for Variational Forms}
The bilinear functions in (\ref{eq:20}) and (\ref{eq:21}) can be explicitly written as
\begin{eqnarray}
c(\textbf{E},\textbf{v})&=&\int_{\Omega}\nabla\times\textbf{v}^{*}\cdot\bar{\bar{\mu}}_{r}^{-1}\nabla\times\textbf{E}dv \label{eq:29}\\
a(\textbf{E},\textbf{v})&=&\int_{\Omega}\textbf{v}^{*}\cdot\bar{\bar{\epsilon}}_{r}\textbf{E}dv \label{eq:30}\\
s_{0}(\textbf{E},\textbf{v})&=&\int_{S_{L}}\textbf{v}^{*}\cdot\hat{n}\times\bar{\bar{\mu}}_{r}^{-1}\nabla\times\textbf{E}d s \label{eq:31}\\
s_{p}(\textbf{H},\textbf{v})&=&\int_{S_{p}}\textbf{v}^{*}\cdot\hat{w}\times\textbf{H}d s \label{eq:32}\\
m(\textbf{a},\textbf{b}) &=& \int_{S_{p}}(\hat{w}\times\textbf{a})^{*}\cdot\hat{w}\times\textbf{b}ds \label{eq:33}
\end{eqnarray}
where $S_{p}$ is the interface ($w=w_{0}$) between the 3-D scattering region $\Omega$ and the inhomogeneous layered media $LM_{p}$, connecting the $p$-th port shown in Fig.\ref{sketch0};
$\hat{n}$ and $\hat{w}$ are the unit outward normal vectors of $\Omega$ and $\hat{n}=\hat{w}$ for the interface $S_{p}$ ;
$S_{L}=\partial\Omega/S_p$ is the rest of the boundary $\partial\Omega$ except $S_{p}$; the symbol ``*'' denotes the complex conjugate of the vector.
Obviously, when $S_L$ is set as PEC/PMC, the boundary integral $s_0$ vanish. Besides, the inner product $(\cdot,\cdot)$ for arbitrary two vectors is defined by
\begin{eqnarray}\label{eq:34}
(\textbf{a},\textbf{b}) = \int_{\Omega}\textbf{b}^{*}\cdot\textbf{a}dv
\end{eqnarray}

\subsection{Matrices Associated with HNM}

First, substituting (\ref{eq:25}) into (\ref{eq:29})-(\ref{eq:31}) and taking the test functions $\textbf{v}=\{\bm{\Phi}_{\alpha}^{c},\nabla\phi_{\xi}\}\in \textbf{W}_{h}$, we can obtain
\begin{equation}\label{eq:35}
\bar{\bar{C}}=
\begin{bmatrix}
\bar{\bar{C}}_{11}&\bar{\bar{0}}\\
\bar{\bar{0}}&\bar{\bar{0}}
\end{bmatrix},
\bar{\bar{A}}=
\begin{bmatrix}
\bar{\bar{A}}_{11}&\bar{\bar{A}}_{12}\\
\bar{\bar{A}}_{21}&\bar{\bar{A}}_{22}
\end{bmatrix},
\bar{\bar{S}}_{0}=
\begin{bmatrix}
\bar{\bar{S}}_{11}&\bar{\bar{0}}\\
\bar{\bar{S}}_{21}&\bar{\bar{0}}
\end{bmatrix}
\end{equation}
where $\bar{\bar{0}}$ denotes the null matrix arise from the vector identity $\nabla\times\nabla\phi=\textbf{0}$. The elements of above block matrices can be written as
\begin{equation*}
\begin{split}
[\bar{\bar{C}}_{11}]_{\alpha,\beta}^{(\kappa)}&=\int_{\kappa}(\nabla\times\bm{\Phi}_{\alpha}^{c})^{*}\cdot\bar{\bar{\mu}}_{r}^{-1}(\nabla_{t}\times\bm{\Phi}_{\beta}^{c})d\kappa\\
[\bar{\bar{A}}_{11}]_{\alpha,\beta}^{(\kappa)}&=\int_{\kappa}(\bm{\Phi}_{\alpha}^{c})^{*}\cdot\bar{\bar{\epsilon}}_{r}\bm{\Phi}_{\beta}^{c}d\kappa\\
[\bar{\bar{A}}_{12}]_{\alpha,\eta}^{(\kappa)}&=\int_{\kappa}(\bm{\Phi}_{\alpha}^{c})^{*}\cdot\bar{\bar{\epsilon}}_{r}\nabla\phi_{\eta}d\kappa\\
[\bar{\bar{A}}_{21}]_{\xi,\beta}^{(\kappa)}&=\int_{\kappa}(\nabla\phi_{\xi})^{*}\cdot\bar{\bar{\epsilon}}_{r}\bm{\Phi}_{\beta}^{c}d\kappa\\
[\bar{\bar{A}}_{22}]_{\xi,\eta}^{(\kappa)}&=\int_{\kappa}(\nabla\phi_{\xi})^{*}\cdot\bar{\bar{\epsilon}}_{r}\nabla\phi_{\eta}d\kappa\\
[\bar{\bar{S}}_{11}]_{\alpha,\beta}^{(\kappa)}&=\int_{\kappa}(\bm{\Phi}_{\alpha}^{c})^{*}\cdot\hat{n}\times\bar{\bar{\mu}}_{r}^{-1}\nabla\times\bm{\Phi}_{\beta}^{c}d\kappa\\
[\bar{\bar{S}}_{21}]_{\xi,\beta}^{(\kappa)}&=\int_{\kappa}(\nabla\phi_{\xi})^{*}\cdot\hat{n}\times\bar{\bar{\mu}}_{r}^{-1}\nabla\times\bm{\Phi}_{\beta}^{c}d\kappa.
\end{split}
\end{equation*}
In the above formulations, $\kappa$ is a physical element of the 3-D scattering region $\Omega$; $\alpha, \beta=1,2,\ldots, N_{c}$ and $\xi, \eta=1,2,\ldots,N_{n}$.
For the compactness, a system matrix $\bar{\bar{K}}$ is defined by $\bar{\bar{K}}=\bar{\bar{C}}-k_{0}^{2}\bar{\bar{A}}+\bar{\bar{S}}_{0}$.

Second, taking $w=w_{0}$ in (\ref{eq:11}) and (\ref{eq:13}), for the incident and scattering fields, from (\ref{eq:32}) the boundary integral $s_{p}$ can be expressed in detail as
\begin{equation}\label{eq:36}
\begin{split}
s_{p}(\textbf{H}_{1,t;p}^{\textrm{scat}},\textbf{v}) = [\int_{S_p}\textbf{v}^{*}\cdot(\bar{\bar{\textbf{H}}}_{1,p}^{+})^{t}ds
+ \int_{S_p}\textbf{v}^{*}\cdot(\bar{\bar{\textbf{H}}}_{1,p}^{-})^{t}ds \bar{\bar{P}}_{1,p}^{-}\tilde{\bar{G}}_{1,2}^{(p)}\bar{\bar{P}}_{1,p}^{+}]\textbf{A}_{1,p}^{\textrm{scat}}
\end{split}
\end{equation}
\begin{equation}\label{eq:37}
s_{p}(\textbf{H}_{1,t;p}^{\textrm{inc}},\textbf{v}) =
\int_{S_p}\textbf{v}^{*}\cdot(\bar{\bar{\textbf{H}}}_{1,p}^{-})^{t}ds \bar{\bar{P}}_{1,p}^{-}\textbf{B}_{1,p}^{\textrm{inc}}
\end{equation}
In view of (\ref{eq:22}), $(\bar{\bar{\textbf{H}}}_{1,p}^{\pm})^{t}$ can be further written as
\begin{equation}\label{eq:38}
\begin{split}
(\bar{\bar{\textbf{H}}}_{1,p}^{\pm})^{t}=\{\bar{\bar{\mathfrak{R}}}\textbf{h}_{1,t}^{\pm(1)},\bar{\bar{\mathfrak{R}}}\textbf{h}_{2,t}^{\pm(1)},\ldots,\bar{\bar{\mathfrak{R}}}\textbf{h}_{m,t}^{\pm(1)}\}
=\{\bar{\bar{\mathfrak{R}}}\textbf{N}_{1},\bar{\bar{\mathfrak{R}}}\textbf{N}_{2},\ldots,\bar{\bar{\mathfrak{R}}}\textbf{N}_{N_{e}}\}\bar{\bar{H}}_{S_{p}}^{\pm}
\end{split}
\end{equation}
where the matrix $\bar{\bar{H}}_{S_{p}}^{\pm}=[\textbf{h}_{1}^{\pm},\textbf{h}_{2}^{\pm},\ldots,\textbf{h}_{m}^{\pm}]$ consisting of the eigenvector $\textbf{h}^{\pm}=[h_{1},h_{2},\ldots,h_{N_{e}}]^{t}$ of the waveguide eigenvalue problem in layer 1 of the layered media $LM_{p}$.
Therefore, taking $\textbf{v}=\{\bm{\Phi}_{\alpha}^{c},\nabla\phi_{\xi}\}\in \textbf{W}_{h}$ and inserting (\ref{eq:38}) into (\ref{eq:36}) and (\ref{eq:37}), we can obtain
\begin{equation}\label{eq:39}
\begin{split}
s_{p}(\textbf{H}_{1,t;p}^{\textrm{scat}},\textbf{v}) = [\bar{\bar{C}}_{S_{p}}\bar{\bar{H}}_{S_{p}}^{+}
+\bar{\bar{C}}_{S_{p}}\bar{\bar{H}}_{S_{p}}^{-} \bar{\bar{P}}_{1,p}^{-}\tilde{\bar{G}}_{1,2}^{(p)}\bar{\bar{P}}_{1,p}^{+}]\textbf{A}_{1,p}^{\textrm{scat}}
\end{split}
\end{equation}
\begin{equation}\label{eq:40}
s_{p}(\textbf{H}_{1,t;p}^{\textrm{inc}},\textbf{v}) =
\bar{\bar{C}}_{S_{p}}\bar{\bar{H}}_{S_{p}}^{-} \bar{\bar{P}}_{1,p}^{-}\textbf{B}_{1,p}^{\textrm{inc}}
\end{equation}
where
\begin{equation*}
\bar{\bar{C}}_{S_{p}}=
\begin{bmatrix}
\bar{\bar{C}}_{S_{p},1}\\
\bar{\bar{C}}_{S_{p},2}
\end{bmatrix}
\end{equation*}
which has the elements as follows
\begin{equation*}
\begin{split}
[\bar{\bar{C}}_{S_{p},1}]_{\alpha,\gamma}^{(\kappa)}&=\int_{\partial\kappa|_{S_{p}}}(\bm{\Phi}_{\alpha}^{c})^{*}\cdot\bar{\bar{\mathfrak{R}}}\textbf{N}_{\gamma}ds\\
[\bar{\bar{C}}_{S_{p},2}]_{\xi,\gamma}^{(\kappa)}&=\int_{\partial\kappa|_{S_{p}}}(\nabla\phi_{\xi})^{*}\cdot\bar{\bar{\mathfrak{R}}}\textbf{N}_{\gamma}ds.
\end{split}
\end{equation*}
Thus, from (\ref{eq:39}) and (\ref{eq:40}), we can define the source vector $\textbf{b}_{S_{p}}=\bar{\bar{C}}_{S_{p}}\bar{\bar{H}}_{S_{p}}^{-} \bar{\bar{P}}_{1,p}^{-}\textbf{B}_{1,p}^{\textrm{inc}}$ and the system matrix $\tilde{\bar{C}}_{S_{p}}=\bar{\bar{C}}_{S_{p}}[\bar{\bar{H}}_{S_{p}}^{+}
+\bar{\bar{H}}_{S_{p}}^{-} \bar{\bar{P}}_{1,p}^{-}\tilde{\bar{G}}_{1,2}^{(p)}\bar{\bar{P}}_{1,p}^{+}]$.

Finally, to derive the system matrices associated with (\ref{eq:21}), $(\bar{\bar{\textbf{F}}}_{1,p}^{\pm})^{t}$ is written as
\begin{equation}\label{eq:41}
\begin{split}
(\bar{\bar{\textbf{F}}}_{1,p}^{\pm})^{t}&=\{\textbf{e}_{1,t}^{\pm(1)},\textbf{e}_{2,t}^{\pm(1)},\ldots,\textbf{e}_{m,t}^{\pm(1)}\}\\
&=\{\textbf{N}_{1},\textbf{N}_{2},\ldots,\textbf{N}_{N_{e}}\}\bar{\bar{E}}_{S_{p}}^{\pm}
\end{split}
\end{equation}
where the matrix $\bar{\bar{E}}_{S_{p}}^{\pm}=[\textbf{e}_{1}^{\pm},\textbf{e}_{2}^{\pm},\ldots,\textbf{e}_{m}^{\pm}]$ consisting of the eigenvector $\textbf{e}^{\pm}=[u_{1},u_{2},\ldots,u_{N_{e}}]^{t}$ of the waveguide eigenvalue problem in layer 1 of $LM_{p}$.
Therefore, substituting (\ref{eq:22}) and (\ref{eq:41}) into (\ref{eq:21}), and noting (\ref{eq:33}), we achieve
\begin{equation}\label{eq:42}
\begin{split}
\bar{\bar{Y}}_{S_{p}}\textbf{e}_{S_{p}} &- \bar{\bar{U}}_{S_{p}}\textbf{A}_{1,p}^{\textrm{scat}}= \textbf{b}_{S_{p}}^{'}
\end{split}
\end{equation}
where
\begin{equation*}
\begin{split}
\bar{\bar{Y}}_{S_{p}}&=(\bar{\bar{E}}_{S_{p}}^{+})^{\dagger}\bar{\bar{M}}_{S_{p}}^{(1)}\\
\bar{\bar{U}}_{S_{p}}&=(\bar{\bar{E}}_{S_{p}}^{+})^{\dagger}\bar{\bar{M}}_{S_{p}}^{(2)}[\bar{\bar{E}}_{S_{p}}^{+}
+\bar{\bar{E}}_{S_{p}}^{-}\bar{\bar{P}}_{1,p}^{-}\tilde{\bar{G}}_{1,2}^{(p)}\bar{\bar{P}}_{1,p}^{+}]\\
\textbf{b}'_{S_{p}}&=(\bar{\bar{E}}_{S_{p}}^{+})^{\dagger}\bar{\bar{M}}_{S_{p}}^{(2)}\bar{\bar{E}}_{S_{p}}^{-}\bar{\bar{P}}_{1,,p}^{-}\textbf{B}_{1}^{\textrm{inc}}\\
\bar{\bar{M}}_{S_{p}}^{(1)}&=
\begin{bmatrix}
\bar{\bar{M}}_{S_{p},1}^{(1)},
\bar{\bar{M}}_{S_{p},2}^{(1)}
\end{bmatrix}.
\end{split}
\end{equation*}
The elements of the matrices $\bar{\bar{M}}_{S_{p}}^{(1)}$ and $\bar{\bar{M}}_{S_{p}}^{(2)}$ are as follows
\begin{equation*}
\begin{split}
[\bar{\bar{M}}_{S_{p},1}^{(1)}]_{\zeta,\beta}^{(\kappa)}&=\int_{\partial\kappa|_{S_{p}}}\hat{w}\times\textbf{N}_{\zeta}\cdot\hat{w}\times \bm{\Phi}_{\beta}^{c}ds\\
[\bar{\bar{M}}_{S_{p},2}^{(1)}]_{\zeta,\eta}^{(\kappa)}&=\int_{\partial\kappa|_{S_{p}}}\hat{w}\times\textbf{N}_{\zeta}\cdot \hat{w}\times \nabla \phi_{\eta}ds\\
[\bar{\bar{M}}_{S_{p}}^{(2)}]_{\zeta,\gamma}^{(\kappa)}&=\int_{\partial\kappa|_{S_{p}}}\hat{w}\times\textbf{N}_{\zeta}\cdot \hat{w}\times \textbf{N}_{\gamma}ds
\end{split}
\end{equation*}
where $\zeta,\gamma=1,2,\ldots,N_{e}$. For the source term, from (\ref{eq:33}), we have
\begin{equation*}
\begin{split}
\textbf{s} = [\int_{\kappa}(\bm{\Phi}_{\alpha}^{c})^{*}\cdot\textbf{S}d\kappa,\int_{\kappa}(\nabla\phi_{\xi})^{*}\cdot\textbf{S}d\kappa]^{t}.
\end{split}
\end{equation*}
Noting that when the $p$-th port is excited by the eigenmodes, the source term $\textbf{s}$ vanish.

\ack
 This research is partially supported by the National Natural Science Foundation of China under Grant 62561013 and by the National Key Research and Development Program of China under Grant 2023YFB3002603.

\end{document}

%% file: latex_defs.tex
\def\^{\hat}
\def\~{\tilde}

\def\3h{{3\over 2}}

\def\eqn#1$${\eqno{{\rm #1}}$$}

\def\~{\tilde}

\def\^{\hat}

\def\bm#1{\mbox{{\boldmath${#1}$}}}
\def\XXint#1#2#3{{\setbox0=\hbox{$#1{#2#3}{\int}$}
     \vcenter{\hbox{$#2#3$}}\kern-.5\wd0}}